\documentclass[12pt,letterpaper]{article}
\usepackage{epsfig,rotating,setspace,latexsym,amsmath,epsf,amssymb,amsfonts,bm,theorem,cite,caption,subcaption,enumerate,longtable,accents}
\usepackage{algorithm,algorithmic,graphicx,epsf,authblk,epstopdf,url,color,multirow}
\newtheorem{theorem}{Theorem}

\newtheorem{corollary}{Corollary}

\newtheorem{remark}{Remark}
\newtheorem{lemma}{Lemma}
\newenvironment{Proof}[1]{\medskip\par\noindent{\bf Proof:\,}\,#1}{{\mbox{\,$\blacksquare$}\par}}

\newcommand{\cs}{{\mathcal{S}}}
\newcommand{\bmu}{{\boldsymbol{\mu}}}
\newcommand{\bt}{{\boldsymbol{\tau}}}

\newcommand{\mds}{{\text{\textbf{MDS}}}}
\newcommand{\st}{{\text{s.t.}}}

\allowdisplaybreaks

\begin{document}
	
\title{Private Information Retrieval Through Wiretap Channel II: Privacy Meets Security\thanks{This work was supported by NSF Grants  CNS 13-14733, CCF 14-22111, CNS 15-26608 and CCF 17-13977. A shorter version is submitted to IEEE ISIT 2018.}}
	
\author{Karim Banawan \qquad Sennur Ulukus\\
	\normalsize Department of Electrical and Computer Engineering\\
	\normalsize University of Maryland, College Park, MD 20742 \\
	\normalsize {\it kbanawan@umd.edu} \qquad {\it ulukus@umd.edu}}
	
\maketitle
	
\vspace*{-0.8cm}

\begin{abstract}
    We consider the problem of private information retrieval through wiretap channel II (PIR-WTC-II). In PIR-WTC-II, a user wants to retrieve a single message (file) privately out of $M$ messages, which are stored in $N$ replicated and non-communicating databases. An external eavesdropper observes a fraction $\mu_n$ (of its choice) of the traffic exchanged between the $n$th database and the user. In addition to the privacy constraint, the databases should encode the returned answer strings such that the eavesdropper learns absolutely nothing about the \emph{contents} of the databases. We aim at characterizing the capacity of the PIR-WTC-II under the combined privacy and security constraints. We obtain a general upper bound for the problem in the form of a max-min optimization problem, which extends the converse proof of the PIR problem under asymmetric traffic constraints. We propose an achievability scheme that satisfies the security constraint by encoding a secret key, which is generated securely at each database, into an artificial noise vector using an MDS code. The user and the databases operate at one of the corner points of the achievable scheme for the PIR under asymmetric traffic constraints such that the retrieval rate is maximized under the imposed security constraint. The upper bound and the lower bound match for the case of $M=2$ and $M=3$ messages, for any $N$, and any $\boldsymbol{\mu}=(\mu_1, \cdots, \mu_N)$.    
\end{abstract}
\section{Introduction}

Private information retrieval (PIR) is a canonical problem which considers the privacy of the content downloaded from public databases. The problem is introduced by Chor et al. \cite{ChorPIR}, and attracted considerable interest within the computer science community \cite{ChorPIR, PIRsurvey2004, cachin1999computationally, ostrovsky2007survey, yekhanin2010private}. In the classical PIR model, there are $N$ replicated and non-colluding databases, each storing the same set of $M$ messages. A user requests to download a single file from the databases privately, i.e., no database can know the identity of the user's desired file. To that end, the user submits a query to each database that does not leak any information about the identity of the file. Each database responds with an answering string. From all answering strings, the user should be able to decode the desired file reliably. PIR schemes are designed to be more efficient than the trivial scheme of downloading all the files stored in the databases. The efficiency is measured by the retrieval rate, which is the ratio between the number of desired message symbols to the total number of downloaded symbols.  PIR is important from a practical point of view as many privacy threats exist in modern networks, in particular, when advanced learning algorithms are employed within social networks and online shopping websites. From a technical standpoint, PIR lies at the intersection of computer science, information theory, coding theory, network coding, and signal processing. 

There has been a growing interest in the PIR problem in the information-theory society, with early examples \cite{RamchandranPIR, unsynchonizedPIR, YamamotoPIR, VardyConf2015, RazanPIR, JafarConf2016}. In \cite{JafarPIR}, Sun and Jafar investigate the fundamental limits of the classical PIR problem by introducing the notion of PIR capacity. The PIR capacity is defined as the supremum of PIR rates over all achievable retrieval schemes. \cite{JafarPIR} determines the exact PIR capacity of the classical model to be $C=(1+\frac{1}{N}+\frac{1}{N^2}+\cdots+\frac{1}{N^{M-1}})^{-1}$. Following \cite{JafarPIR}, the fundamental limits of many interesting variants of the classical PIR problem have been considered, such as: PIR from colluding databases, robust PIR, symmetric PIR, PIR from MDS-coded databases, PIR for arbitrary message lengths, multi-round PIR, multi-message PIR, PIR from Byzantine databases, secure symmetric PIR with adversaries, cache-aided PIR, PIR with private side information (PSI), PIR for functions, storage constrained PIR, PIR with asymmetric traffic constraints and their several combinations 
\cite{JafarColluding,symmetricPIR,KarimCoded,arbmsgPIR,codedsymmetric,MultiroundPIR,codedcolluded,codedcolludedJafar,arbitraryCollusion,MPIRjournal,codedcolludingZhang,MPIRcodedcolludingZhang,BPIRjournal,tandon2017capacity,wang2017linear,kadhe2017private,wei2017fundamental,chen2017capacity,wei2017capacity,sun2017_computation,mirmohseni2017private,abdul2017private,wei2017fundamental_partial,subpacketization_capacity,PFR_colluding,SecureSymmetricPIR,SecurePIR,KarimAsymmetricPIR}.

The sole requirement of most of these previous works is to protect the identity of the desired message from the public databases in addition to satisfying the reliability constraint. We ensure this protection via imposing the privacy constraint on the submitted queries. Another interesting dimension to the PIR problem is when the content of the requested message needs to be protected against an external eavesdropper (wiretapper), who wishes to learn about the contents of the databases by observing the queries and answer strings exchanged between the user and the databases. In this paper, we tackle the problem of secure PIR. We impose an extra constraint to the PIR problem, namely, the secrecy constraint in addition to the usual privacy constraint. The secrecy constraint ensures that the queries and the answer strings do not leak any information about the contents of the databases to the eavesdropper. Such systems are relevant in practice, for example, in the stock market, investors need to keep the identity of the records that they are interested in private from the public databases as revealing such interest in a specific record may change its value. This is a classical PIR application. Now, consider the case when the contents of the records themselves are confidential except for a small subset of authorized investors. Thus, the queries and the answer strings should be designed such that unauthorized entities who wiretap the retrieval process learn absolutely nothing about the contents of these confidential records.  

Although there is a vast literature on PIR, only a few works exist on secure PIR: \cite{SSRI} considers the more general problem of information storage and retrieval, guaranteeing that also the process of storing the information is secure in the presence of failing servers. \cite{SecureSymmetricPIR} considers a \emph{symmetric} PIR setting where there is a passive eavesdropper who can tap in on the incoming and outgoing transmissions of any $E$ servers. \cite{SecureSymmetricPIR} derives the PIR capacity in this setting. Interestingly, the secret key needed for the \emph{symmetric} retrieval process is used as an encryption key to secure the contents of the databases from the eavesdropper. This requires, as in the underlying symmetric PIR, that databases exchange a secret key of at least a certain size. This problem is investigated further in \cite{SecurePIR} for the classical PIR problem under $T$-privacy constraint for the case of $E \leq T$. \cite{SecurePIR} derives inner and outer bounds for this problem in addition to the minimum amount of common randomness required, which is shared between the databases.

We study the secure PIR problem from a different angle than \cite{SSRI,SecureSymmetricPIR,SecurePIR}. We consider a classical PIR setting, where there are $N$ replicated databases storing $M$ messages. We assume that the contents of the databases are fixed and cannot be coded to satisfy the security constraint during the storage phase, unlike \cite{SSRI}. There are no shared keys in place required for symmetric PIR unlike \cite{SecureSymmetricPIR}, as we consider classical PIR, not symmetric PIR. We further assume that the eavesdropper observes the queries and the answer strings of all databases through wiretap channels in contrast to observing the noiseless transmission from any $E$ of the databases as in \cite{SecurePIR}. In this work, we investigate the PIR problem through wiretap channel II (PIR-WTC-II). Ozarow and Wyner \cite{WTC-II} introduced the wiretap channel II (WTC-II) model, which considers a noiseless main channel and a binary erasure channel to the wiretapper, where the wiretapper is able to select the positions of erasures. In PIR-WTC-II (see Fig.~\ref{PIR_WTC_II}), the user observes the $t_n$-length answer strings through a noiseless channel from the $n$th database. The eavesdropper can observe a fraction $\mu_n$ from the $n$th answer string. More specifically, the eavesdropper chooses any set of positions $\mathcal{S}_n \subset \{1, \cdots, t_n\}$ to observe from the $n$th answer string, such that $|\mathcal{S}_n|=\mu_n t_n$. The databases should encode the answer strings such that the eavesdropper learns nothing from observing any $\mu_n$ fraction of the traffic from the $n$th database. This is in addition to normal privacy and reliability constraints. Naturally, the $n$th database dedicates $\mu_n t_n$ portion of the answer string to confuse the eavesdropper, constraining the \emph{meaningful} portion of the answer to be $(1-\mu_n)t_n$. This fundamentally relates PIR-WTC-II to the PIR problem under asymmetric traffic constraints \cite{KarimAsymmetricPIR}, as lengths of answer strings can no longer be symmetric. This poses the following questions: How can we design a retrieval code that satisfies the combined privacy and security constraints for the PIR-WTC-II problem? Does PIR-WTC-II problem necessitate the existence of common randomness between the databases as in \cite{SecurePIR}? Should the databases share any common randomness with the user (retriever)?  

In this paper, we obtain a general upper bound for the PIR-WTC-II problem, when the eavesdropper can wiretap $\bmu=(\mu_1, \cdots, \mu_N)$ fractions from the traffic outgoing from every database. We note first that this problem is the first concrete example of a PIR problem under asymmetric traffic constraints in the sense of \cite{KarimAsymmetricPIR}. We show that this upper bound can be expressed as a max-min problem. The inner minimization problem extends the converse techniques of the PIR problem under asymmetric traffic constraints in \cite{KarimAsymmetricPIR} to the PIR-WTC-II problem. The outer problem maximizes the retrieval rate over all possible traffic ratio vectors. For the achievability, we extend the achievable scheme used in \cite{KarimAsymmetricPIR} to achieve the corner points for the \emph{meaningful} portions of the queries. In the extension, to satisfy the security constraint, each database generates a secret key with $\mu_n t_n$ length and encodes it into an artificial noise vector using a $(t_n, \mu_n t_n)$ MDS code and encrypts the returned answer strings with the artificial noise vector. Interestingly, our achievable rate does not need any shared randomness among the databases or between the databases and the user. The keys used by the databases are unknown to the user, but are decodable and canceled at the retriever; however, the same keys are not extractable at the wiretapper due to the MDS code used and the existence of WTC-II. We express the achievable retrieval rate in terms of the output of a system of difference equations. We present an explicit achievable rate for the problem for the case of $N=2$ databases and any arbitrary $M$. Our upper and lower bounds match for $M=2$ and $M=3$ messages, for any $N$, and any $\bmu$, which conforms with the results of \cite{KarimAsymmetricPIR}.   
\section{System Model}
Consider a classical PIR model, in which there are $N$ non-colluding and replicated databases, each storing the same content of $M$ messages (or files). The message $W_m$ is represented as a vector of length $L$, whose elements are picked from a finite field $\mathbb{F}_q^L$ with a sufficiently large alphabet. The messages $W_{1:M}=\{W_1, \cdots, W_M\}$ are independent and identically distributed, hence,
\begin{align}
H(W_m)&=L, \quad m \in \{1, \cdots,M\}\\
H(W_{1:M})&=ML, \quad (q\text{-ary bits})
\end{align}
We assume that the messages are uncoded and fixed, i.e., we assume that the contents of the databases cannot be coded to satisfy the security constraint during the storage phase.

In classical PIR, a user wants to retrieve a message $W_i$ from the $N$ databases without revealing the identity of the message $i$ to any individual database. The user prepares $N$ queries, one for each database. The user sends $Q_n^{[i]}$ to the $n$th database. Since the user has no knowledge about the realization of $W_{1:M}$, the queries and the messages are statistically independent, i.e.,
\begin{align}\label{independency}
I(Q_{1:N}^{[i]};W_{1:M})=0, \quad i \in \{1, \cdots, M\}
\end{align}
where $Q_{1:N}^{[i]}=\{Q_1^{[i]}, \cdots, Q_N^{[i]}\}$. Furthermore, to ensure the privacy of $W_i$, the user should constrain the query intended to retrieve $W_i$ to be indistinguishable from the query intended to retrieve any other message $W_{j}$ at any individual database. Thus, the privacy constraint is formalized as,
\begin{align}\label{privacy_constraint}
(Q_n^{[i]}, A_n^{[i]}, W_{1:M}) \sim (Q_n^{[j]}, A_n^{[j]}, W_{1:M}), \quad \forall j \in \{1, \cdots, M\}
\end{align} 
where $\sim$ denotes statistical equivalence.

The $n$th database, after receiving the query $Q_n^{[i]}$, responds with a $t_n$-length answering string $A_n^{[i]}$. Note that we allow the user and the databases to choose arbitrary lengths for the answer strings such that they maximize the retrieval rate. The answer string is generally a \emph{stochastic} mapping of the messages $W_{1:M}$ and the received query $Q_n^{[i]}$, hence,
\begin{align}\label{answer_constraint}
H(A_n^{[i]}|Q_n^{[i]}, W_{1:M},\mathcal{G}_n)=0, \quad n \in \{1, \cdots, N\}
\end{align}
where $\mathcal{G}_n$ is a random variable independent of all other random variables, whose realization is known at the $n$th database only and not shared with any other database or the user a priori of the transmission. We denote the traffic ratio vector by $\boldsymbol{\tau}=(\tau_1, \cdots, \tau_N)$. The traffic ratio at the $n$th database $\tau_n$ is given by,
\begin{align}
\tau_n=\frac{t_n}{\sum_{i=1}^{N} t_i}
\end{align}

\begin{figure}[t]
	\centering
	\includegraphics[width=0.8\textwidth]{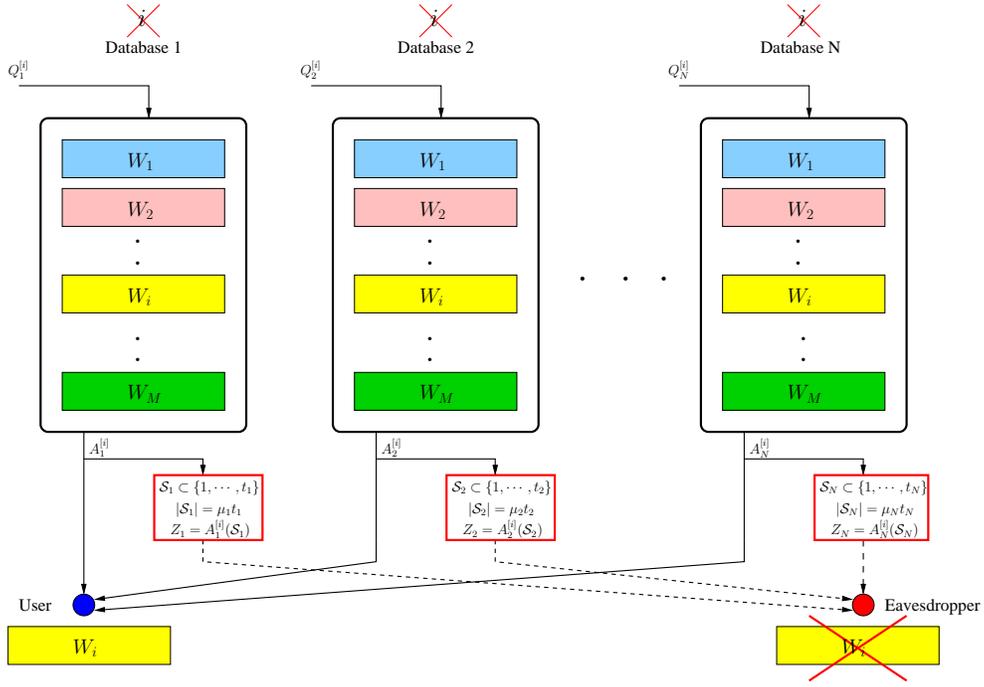}
	\caption{Secure PIR problem through wiretap channel II.}
	\label{PIR_WTC_II}
	\vspace*{-0.4cm}
\end{figure}

We assume that the answer strings are transmitted through a WTC-II (see Fig.~\ref{PIR_WTC_II}). In this case, an external eavesdropper (wiretapper) wishes to learn about the contents of the databases by observing the queries and answer strings exchanged by the user and the databases. In PIR-WTC-II, the user observes the $t_n$-length answer string $A_n^{[i]}$ from the $n$th database through a noiseless channel. On the other hand, the eavesdropper can observe a fraction $\mu_n$ from the $n$th answer string. More specifically, the eavesdropper arbitrarily chooses any set of positions $\mathcal{S}_n \subset \{1, \cdots, t_n\}$ to observe from the $n$th answer string, such that $|\mathcal{S}_n|=\mu_n t_n$, i.e., the output of the eavesdropper channel is given by,
\begin{align}
Z_n^{[i]}=A_n^{[i]}(\mathcal{S}_n), \quad n \in \{1, \cdots, N\}
\end{align} 
We denote the unobserved portion of the answer string by $Y_n^{[i]}=A_n^{[i]}(\bar{\mathcal{S}}_n)$, where $\bar{\mathcal{S}}_n=\{1, \cdots, N\}\setminus \mathcal{S}_n$, thus, $A_n^{[i]}=(Y_n^{[i]},Z_n^{[i]})$. We write the eavesdropping ratios as a vector $\boldsymbol{\mu}=(\mu_1, \cdots, \mu_N)$. Without loss of generality, we assume that the databases are arranged ascendingly in $\mu_n$, i.e., $\mu_1 \leq \mu_2 \leq \cdots \leq \mu_N$, i.e., the first database is the least threatened (most secure) and the $N$th database is the most threatened (least secure).

Upon preparing the answer string, the databases should encode the answer strings such that the eavesdropper learns nothing from observing any $\mu_n$ fraction from the traffic from the $n$th database even with observing the queries submitted by the user. Consequently, we write the security constraint as,
\begin{align}\label{security_constraint}
I(W_{1:M};Z_{1:N}^{[i]},Q_{1:N}^{[i]})=0
\end{align} 

Additionally, the user should be able to reconstruct the desired message $W_i$ from the collected answer strings with arbitrarily small probability of error. Using Fano's inequality, we write the reliability constraint as,
\begin{align}\label{reliability_constraint}
	H(W_i|Q_{1:N}^{[i]},A_{1:N}^{[i]})=o(L)
\end{align}
where $\frac{o(L)}{L} \rightarrow 0$ as $L \rightarrow \infty$. 

For a fixed $N$, $M$, traffic ratio vector $\bt$, and eavesdropping ratio vector $\boldsymbol{\mu}$, a retrieval rate $R(\boldsymbol{\tau},\boldsymbol{\mu})$ is achievable if there exists a PIR scheme which satisfies the privacy constraint \eqref{privacy_constraint}, security constraint \eqref{security_constraint}, and the reliability constraint \eqref{reliability_constraint} for some message length $L(\bt,\boldsymbol{\mu})$ and answer strings of lengths $\{t_n(\bt, \boldsymbol{\mu})\}_{n=1}^N$ such that $\tau_n=\frac{t_n(\bt, \boldsymbol{\mu})}{\sum_{i=1}^{N} t_i(\bt, \boldsymbol{\mu})}$, where the retrieval rate is therefore given by,  
\begin{align}
R(\bt, \boldsymbol{\mu})=\frac{L(\bt,\boldsymbol{\mu})}{\sum_{n=1}^N t_n(\bt, \boldsymbol{\mu})}
\end{align}
We note that in this problem, the user and the databases can agree on a traffic ratio vector $\boldsymbol{\tau}$ to maximize the retrieval rate, thus, we can express the secure retrieval rate under eavesdropping capabilities $\bmu$, $R(\boldsymbol{\mu})$, as,
\begin{align}
R(\boldsymbol{\mu})=\max_{\boldsymbol{\tau}} \:\: R(\boldsymbol{\tau},\boldsymbol{\mu})
\end{align}
Note that the message lengths can grow arbitrarily large to conform with standard information-theoretic arguments. The capacity of the PIR-WTC-II problem $C(\boldsymbol{\mu})$ is defined as the supremum of all achievable retrieval rates over all achievable schemes, i.e., $C(\boldsymbol{\mu})=\sup \:R(\boldsymbol{\mu})$. 

\section{Main Results and Discussions}
In this section, we present the main results of this paper. Our first result characterizes a general upper bound for the PIR-WTC-II problem for fixed $M$, $N$, and an arbitrary $\bmu$.
\begin{theorem}[Upper bound]\label{Thm1}
	For the PIR-WTC-II problem under eavesdropping capabilities $\bmu=(\mu_1, \cdots,\mu_N)$, the capacity is upper bounded by,
	\begin{align}\label{upper_bound}
	C(\bmu) \leq \bar{C}(\bmu)=\max_{\boldsymbol{\tau} \in \mathbb{T}} \min_{n_i \in \{1, \cdots, N\}} \frac{\sum_{n=1}^N (1-\mu_n)\tau_n+\frac{\sum_{n=n_1+1}^N (1-\mu_n)\tau_n}{n_1}+\cdots+\frac{\sum_{n=n_{M-1}+1}^N (1-\mu_n)\tau_n}{\prod_{i=1}^{M-1} n_i}}{1+\frac{1}{n_1}+\cdots+\frac{1}{\prod_{i=1}^{M-1}n_i}}
	\end{align}
 where $\mathbb{T}=\left\{\boldsymbol{\tau}: \tau_n \geq 0 \quad \forall n \in [1:N],\quad \sum_{n=1}^{N} \tau_n=1\right\}$.
\end{theorem}

The proof of this upper bound is given in Section~\ref{converse}. We have the following remarks.

\begin{remark}
	When $\bmu=(0, \cdots,0)$, i.e., without any security constraints, the upper bound reduces to:
	\begin{align}
	\bar{C}(\bmu)&=\max_{\boldsymbol{\tau} \in \mathbb{T}} \min_{n_i \in \{1, \cdots, N\}} \frac{\sum_{n=1}^N \tau_n+\frac{\sum_{n=n_1+1}^N \tau_n}{n_1}+\cdots+\frac{\sum_{n=n_{M-1}+1}^N \tau_n}{\prod_{i=1}^{M-1} n_i}}{1+\frac{1}{n_1}+\cdots+\frac{1}{\prod_{i=1}^{M-1}n_i}}\\
	&=\max_{\boldsymbol{\tau} \in \mathbb{T}} \min_{n_i \in \{1, \cdots, N\}} \frac{1+\frac{\sum_{n=n_1+1}^N \tau_n}{n_1}+\cdots+\frac{\sum_{n=n_{M-1}+1}^N \tau_n}{\prod_{i=1}^{M-1}\label{security-traffic} n_i}}{1+\frac{1}{n_1}+\cdots+\frac{1}{\prod_{i=1}^{M-1}n_i}}\\
	&=\max_{\boldsymbol{\tau}}\quad \tilde{C}(\bt)\\
	&=\frac{1}{1+\frac{1}{N}+\cdots+\frac{1}{N^{M-1}}}
	\end{align}
	where the inner problem in \eqref{security-traffic} is precisely the upper bound of the PIR problem under asymmetric traffic $\bt$ \cite{KarimAsymmetricPIR}. From \cite{KarimAsymmetricPIR}, we know that $\tilde{C}(\bt)$ is maximized by adopting symmetric schemes, i.e., $\tau_n=\frac{1}{N}$, which achieves the PIR capacity $C$ in \cite{JafarPIR}.
\end{remark}

\begin{remark}
	If the PIR-WTC-II problem is further constrained by the asymmetric traffic constraints $\bt$, the corresponding upper bound $\bar{C}(\bmu,\bt)$ is given by the inner problem of \eqref{upper_bound}, i.e., 
	\begin{align}
	\bar{C}(\bmu,\bt)=\min_{n_i \in \{1, \cdots, N\}} \frac{\sum_{n=1}^N (1-\mu_n)\tau_n+\frac{\sum_{n=n_1+1}^N (1-\mu_n)\tau_n}{n_1}+\cdots+\frac{\sum_{n=n_{M-1}+1}^N (1-\mu_n)\tau_n}{\prod_{i=1}^{M-1} n_i}}{1+\frac{1}{n_1}+\cdots+\frac{1}{\prod_{i=1}^{M-1}n_i}}
	\end{align}
	Hence, without the asymmetric traffic constraints, the user and the databases can agree on $\bt$ that maximizes the retrieval rate, which results in the outer maximization over $\bt$. This is reminiscent of the classical converse proof for the channel coding theorem, where a converse argument is constructed for an arbitrary input distribution of the transmission codebook, and then the converse proof is concluded with a maximization step over all the input distributions.
\end{remark}

\begin{remark}\label{Remark3}
	The upper bound $\bar{C}(\bmu)$ in Theorem \ref{Thm1} can be written as the following linear programming problem:
    \begin{align}
    \bar{C}(\bmu)=\max_{\bt,R}  &\quad R \notag\\
    \st &\quad R \leq \frac{\sum_{n=1}^N (1-\mu_n)\tau_n+\frac{\sum_{n=n_1+1}^N (1-\mu_n)\tau_n}{n_1}+\cdots+\frac{\sum_{n=n_{M-1}+1}^N (1-\mu_n)\tau_n}{\prod_{i=1}^{M-1} n_i}}{1+\frac{1}{n_1}+\cdots+\frac{1}{\prod_{i=1}^{M-1}n_i}}, \quad \forall \mathbf{n}\notag\\
    &\quad \tau_n \geq 0, \quad n=1, \cdots, N \notag\\
    &\quad \sum_{n=1}^{N} \tau_n=1
    \end{align}	
    where $\mathbf{n}=(n_1, \cdots, n_{M-1}) \subset \{1, \cdots, N\}^{M-1}$, i.e., the number of constraints are finite (at most $N^{M-1}+2$ constraints). Hence, the optimal solution of this optimization problem is attained at one of the corner points of the feasible set.  
\end{remark}

Next, we present a general lower bound on $C(\bmu)$ for fixed $M$, $N$.
\begin{theorem}[Lower bound]\label{Thm2}
	For PIR-WTC-II, for a monotone non-decreasing sequence $\mathbf{n}=\{n_i\}_{i=0}^{M-1} \subset \{1, \cdots, N\}^{M}$, let $n_{-1}=0$, and $\cs=\{i \geq 0: n_i-n_{i-1}>0\}$. Denote $y_\ell[k]$ to be the number of stages of the achievable scheme that downloads $k$-sums from the $n$th database in one repetition of the scheme, such that $n_{\ell-1} \leq n \leq n_{\ell}$, and $\ell \in \cs$. Let $\xi_\ell=\prod_{s \in \cs \setminus \{\ell\}} \binom{M-2}{s-1}$. The number of stages $y_\ell[k]$ is characterized by the following system of difference equations:
	\begin{align}\label{differenceEqn}
	y_0[k]&=(n_0\!-\!1)y_0[k\!-\!1]+\sum_{j \in \cs \setminus \{0\}} (n_j\!-\!n_{j-1}) y_j[k\!-\!1] \notag\\
	y_1[k]&=(n_1\!-\!n_0\!-\!1)y_1[k\!-\!1]+\sum_{j \in \cs \setminus \{1\}} (n_j\!-\!n_{j-1}) y_j[k\!-\!1] \notag\\
	y_\ell[k]&=n_0 \xi_\ell \delta[k\!-\!\ell\!-\!1]+(n_\ell\!-\!n_{\ell-1}\!-\!1) y_\ell[k-1]+\sum_{j \in \cs \setminus \{\ell\}} (n_j\!-\!n_{j-1})y_j[k\!-\!1], \quad  \ell \geq 2
	\end{align} 
	where $\delta[\cdot]$ denotes the Kronecker delta function. The initial conditions of \eqref{differenceEqn} are $y_0[1]=\prod_{s \in \cs} \binom{M-2}{s-1}$, and $y_j[k]=0$ for $k \leq j$. Consequently, the traffic ratio vector $\bt(\mathbf{n})=(\tau_1(\mathbf{n}), \cdots, \tau_N(\mathbf{n}))$ corresponding to the sequence $\mathbf{n}=\{n_i\}_{i=0}^{M-1}$ is given by:
	\begin{align}
	\tau_n(\mathbf{n})=\frac{\sum_{k=1}^M \binom{M}{k} y_j[k]}{\sum_{\ell \in \cs} \sum_{k=1}^M \binom{M}{k} y_\ell[k] (n_\ell-n_{\ell-1})},\quad n_{j-1}+1 \leq n \leq n_j
	\end{align}
	Then, the achievable rate corresponding to $\mathbf{n}$ is given by:
\begin{align}
R(\mathbf{n},\bmu)=\frac{\sum_{\ell \in \cs} \sum_{k=1}^{M}\binom{M-1}{k-1} y_\ell[k](n_\ell-n_{\ell-1})}{\sum_{\ell \in \cs} \sum_{n=n_{\ell-1}+1}^{n_\ell}\frac{\sum_{k=1}^{M} \binom{M}{k} y_\ell[k] }{1-\mu_n}}
\end{align}
Consequently, the capacity $C(\bmu)$ is lower bounded by:
\begin{align}
C(\bmu) \geq R(\bmu)&=\max_{n_0 \leq \cdots \leq n_{M-1} \in \{1, \cdots, N\}} R(\mathbf{n},\bmu)\\
 &=\max_{n_0 \leq \cdots \leq n_{M-1} \in \{1, \cdots, N\}} \frac{\sum_{\ell \in \cs} \sum_{k=1}^{M}\binom{M-1}{k-1} y_\ell[k](n_\ell-n_{\ell-1})}{\sum_{\ell \in \cs} \sum_{n=n_{\ell-1}+1}^{n_\ell}\frac{\sum_{k=1}^{M} \binom{M}{k} y_\ell[k] }{1-\mu_n}}
\end{align} 
\end{theorem}

The proof of Theorem \ref{Thm2} can be found in Section~\ref{achievability}. We have the following remarks.

\begin{remark}
	For fixed $M$, $N$, the number of the achievable rates $R(\mathbf{n},\bmu)$ in Theorem~\ref{Thm2} corresponds to the number of monotone non-decreasing sequences $\mathbf{n}=\{n_i\}_{i=0}^{M-1}$, which is equal to $\binom{M+N-1}{M}$.
\end{remark}

\begin{remark}\label{time-sharing}
	After achieving the corner points in Theorem~\ref{Thm2}, which achieve $R(\mathbf{n},\bmu)$, one can perform time-sharing between the corner points to obtain an achievable $R(\bt,\bmu)$ for any $\bt$. The highest possible achievable rate can be obtained by maximizing over $\bt$. However, this is not needed as time-sharing results in a piece-wise affine function in $\bt$. Hence, maximizing over $\bt$ would result in operating directly at one of the corner points. 
\end{remark}

\begin{remark}
	We note that the core of the achievability scheme is the PIR scheme under asymmetric traffic constraints in \cite{KarimAsymmetricPIR}. Hence, the recursive structure described by \eqref{differenceEqn} is directly inherited from \cite{KarimAsymmetricPIR}. Nevertheless, two main differences appear in the final rate expression. First, the answer string length from every database belonging to the same group is different in contrast to \cite{KarimAsymmetricPIR}. This is due to the fact that every database experiences a different eavesdropping capability $\mu_n$ in general, hence the $n$th database encrypts its responses with a key, whose length depends on $\mu_n$, thus the key lengths are different in general. Second, there is no need for time-sharing over the corner points as shown in Remark~\ref{time-sharing}.
\end{remark}

In the following corollary, we settle the capacity $C(\bmu)$ for $M=2$, $M=3$, and arbitrary $N$.
\begin{corollary}[Exact capacity for $M=2$ and $M=3$ messages]\label{M3N2}
	For PIR-WTC-II, the capacity $C(\bmu)$ for $M=2,3$, and an arbitrary $N$ is given by:
		\begin{align}\label{capacityM32}
		C(\bmu) =  
		\left\{
		\begin{array}{ll}
		\max_{n_0,n_1 \in \{1, \cdots, N\}} \frac{n_0 n_1}{\sum_{n=1}^{n_0} \frac{n_0+1}{1-\mu_n}+\sum_{n=n_0+1}^{n_1} \frac{n_0}{1-\mu_n}}, \:\: &M=2 \\
		\max_{n_0,n_1,n_2 \in \{1, \cdots, N\}} \frac{n_0n_1n_2}{\sum_{n=1}^{n_0} \frac{n_0n_1+n_0+1}{1-\mu_n}+\sum_{n=n_0+1}^{n_1} \frac{n_0n_1+n_0}{1-\mu_n}+\sum_{n=n_1+1}^{n_2} \frac{n_0n_1}{1-\mu_n}}, \:\: &M=3
		\end{array}
		\right.
		\end{align}
\end{corollary}

The proof of Corollary~\ref{M3N2} can be found in Section~\ref{proofM3N2}. 

\begin{figure}[t]
	\centering
	\includegraphics[width=0.7\textwidth]{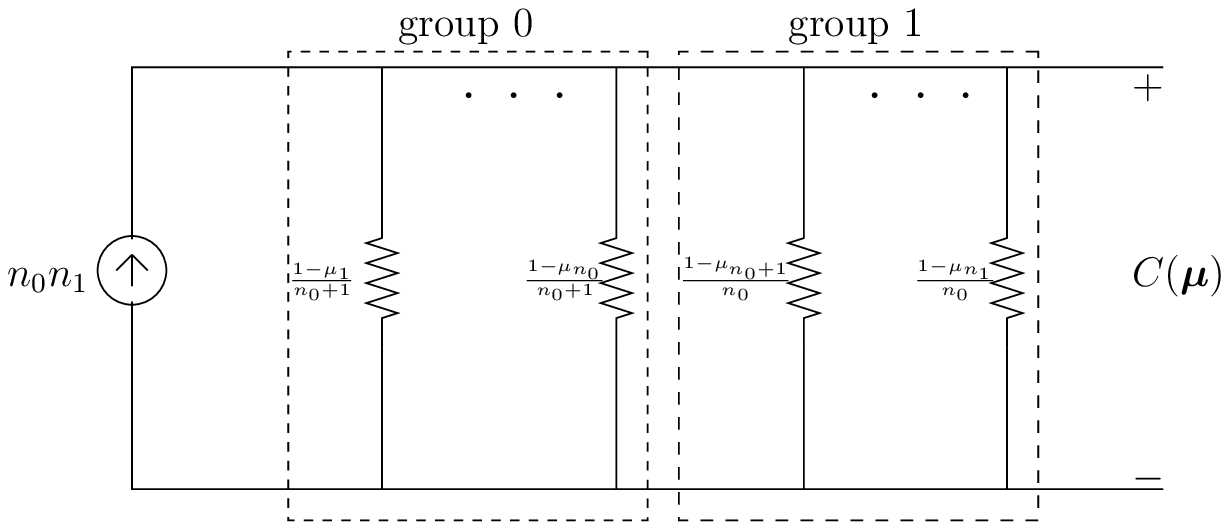}
	\caption{Circuit interpretation of $C(\bmu)$ for $M=2$.}
	\label{Fig:circuit M2}
	\vspace*{-0.4cm}
\end{figure}

\begin{figure}[t]
	\centering
	\includegraphics[width=1\textwidth]{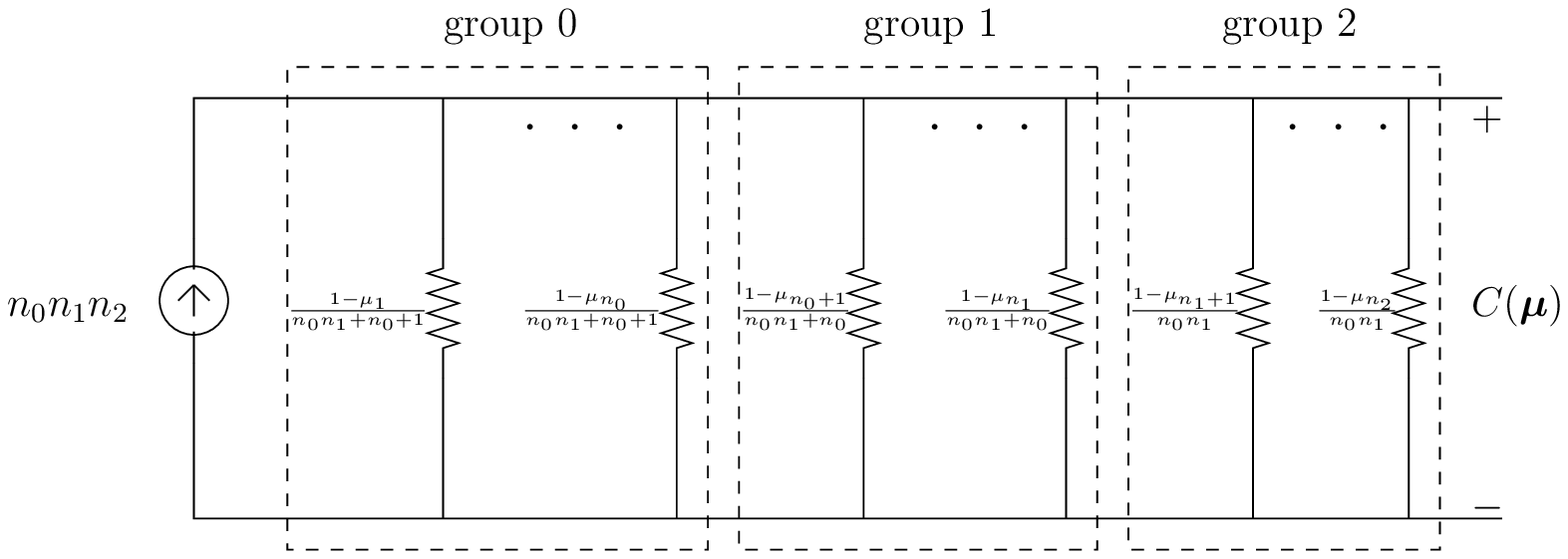}
	\caption{Circuit interpretation of $C(\bmu)$ for $M=3$.}
	\label{Fig:circuit M3}
	\vspace*{-0.4cm}
\end{figure}

\begin{remark}
	The explicit capacity expressions in Corollary~\ref{M3N2} can be interpreted using basic circuit theory. To see that for $M=2$ for a given $(n_0,n_1)$, consider the circuit in Fig.~\ref{Fig:circuit M2}. The circuit has a current source of $n_0 n_1$ units. The circuit consists of $n_0+n_1$ parallel resistors. The $n$th resistor has the value of $R_n=\frac{1-\mu_n}{n_0+1}$ if $1 \leq n \leq n_0$, and $R_n=\frac{1-\mu_n}{n_0}$ if $n_0+1 \leq n \leq n_1$. Hence, the capacity $C(\bmu)$ is the voltage across the current source. A similar interpretation can be inferred from Fig.~\ref{Fig:circuit M3} for the case of $M=3$. Interestingly, this interpretation implies that in order to maximize the retrieval rate (the voltage across the equivalent resistance of the circuit), one should pick $n_0, n_1, n_2$ such that the resistance of each parallel branch is as symmetric as possible. This is due to the fact that the equivalent resistance of parallel resistors is less than the resistance of the least resistor. 
\end{remark}
  
		
Finally, in the next corollary, we present an explicit achievable rate for $R(\bmu)$ when $N=2$, and an arbitrary $M$. The proof of the corollary can be found in Section~\ref{N2}

\begin{corollary}[Achievable retrieval rate for $N=2$]
	For PIR-WTC-II with $N=2$ and an arbitrary $M$, let $s_2=\{1, \cdots, M-1\}$, then the secure PIR capacity $C(\bmu)$ is lower bounded by:
	\begin{align}\label{achievableN2}
	\max\left\{\frac{1-\mu_1}{M},
	                        \!\max_{s_2 \in \{0, \cdots, M-1\}}\frac{\binom{M-2}{s_2-1}+\sum_{k=0}^{M-s_2-1}\binom{M-1}{s_2+k}}{\frac{1}{1-\mu_1} \left[M\binom{M-2}{s_2-1}
		\!\!+\!\!\sum_{k=1}^{\left\lfloor \frac{M-s_2}{2}\right\rfloor} \binom{M}{s_2+2k} \right]\!+\!\frac{1}{1-\mu_2}\!\left[\sum_{k=0}^{\left\lfloor \frac{M-s_2-1}{2}\right\rfloor} \binom{M}{s_2+2k+1}\right]}\right\}
	\end{align}
\end{corollary}

\begin{remark}
	We note the strong connection between the PIR-WTC-II problem and the PIR problem under asymmetric traffic constraints in \cite{KarimAsymmetricPIR}. In PIR-WTC-II problem, the $n$th database uses a secret key of length $\mu_n t_n$ to span the entire space of the eavesdropper. This in turn leaves $(1-\mu_n)t_n$ symbols for meaningful queries. Since the eavesdropping vulnerabilities of the databases are different in general (different $\mu_n$), the meaningful queries are naturally constrained, e.g., we expect the first database (the most secure) to support more meaningful queries than the remaining databases. However, the main difference between the two problems is that in the PIR problem under asymmetric traffic constraints \cite{KarimAsymmetricPIR}, the traffic ratio vector $\bt$ is fixed (by the problem formulation) in contrast to the PIR-WTC-II problem, where the user and the databases can agree on a traffic ratio vector $\bt$ to maximize the retrieval rate under the fixed eavesdropping capabilities $\bmu$.
\end{remark}

\begin{remark}
	
	We now compare our model with the PIR model in \cite{SecureSymmetricPIR, SecurePIR}. In \cite{SecureSymmetricPIR, SecurePIR}, there is an eavesdropper, which observes all communication of $E$ out of $N$ databases, whose identities are unknown to the user.  We restrict the comparison to the case $T=1$ (i.e., no collusion between the databases). In this case, the capacity of the secure PIR problem in \cite{SecurePIR} (abbreviated as T-EPIR problem) is $1-\frac{E}{N}$. This requires a common randomness, which is shared between the databases and unknown to the user, of length $\frac{E}{N-E}$ \cite[Theorem~1]{SecurePIR}. We note that the capacity expression is independent of the number of messages in \cite{SecurePIR}. For the symmetric version of the problem in \cite{SecureSymmetricPIR}, the capacity expression is also $1-\frac{E}{N}$. Interestingly, in the symmetric version of the problem, the common randomness among the databases is used to satisfy both the database privacy and the security constraints simultaneously. 
	
	On the other hand, in our model, the eavesdropper wiretaps all $N$ databases according to the given $\bmu=(\mu_1, \cdots, \mu_N)$. The user knows the ratio of the traffic which is observed by the eavesdropper from each database, i.e., $\bmu=(\mu_1, \cdots, \mu_N)$, but does not know which positions are being observed. Surprisingly, our model does not need any shared randomness among the databases or with the user, i.e., here we are able to achieve nontrivial PIR rates with zero shared randomness rates.
	
	As a concrete example, let $M=3$, and for a fair comparison, let $\mu_n=\frac{E}{N}$ for all $n \in \{1, \cdots, N\}$ in our model. The rationale for this choice of $\mu_n$ is that in \cite{SecurePIR}, the eavesdropper has access to a total of $E\cdot t$ observations, where $t$ is the length of the answer string from any database in \cite{SecurePIR}. Now, for symmetric $\mu_n=\frac{E}{N}$ in our model, all answer string lengths need to be symmetric, i.e., $t_n=t$ for all $n$, and therefore, the eavesdropper accesses a total of $\frac{E}{N} \cdot N \cdot t = E \cdot t$ observations here as it does in \cite{SecurePIR}. The capacity for this case in our model, from Corollary~\ref{M3N2}, is $\frac{1-\frac{E}{N}}{1+\frac{1}{N}+\frac{1}{N^2}}$, which is attained with $n_0=n_1=n_2=N$ in the corollary. This rate is strictly less than the rate in \cite{SecurePIR}, which is $1-\frac{E}{N}$, however, \cite{SecurePIR} requires a shared randomness between the databases at a rate of at least $\frac{E}{N-E}$, while in our case no shared randomness is required.
	
\end{remark}
  
\section{Converse Proof}\label{converse}
In this section, we derive a general upper bound for the retrieval rate under the privacy and security constraints \eqref{privacy_constraint}, \eqref{security_constraint} for the PIR-WTC-II problem. Our converse proof extends the techniques of \cite{JafarPIR} to incorporate the security constraint. In addition, since the eavesdropper observes a different fraction of the traffic from each database, we do not expect that the answer strings (and consequently the traffic ratios) from each database to be symmetric in length. Thus, we modify the converse proof in \cite{JafarPIR} to account for this prospected traffic asymmetry along the lines of \cite{KarimAsymmetricPIR}. However, different from \cite{KarimAsymmetricPIR}, traffic ratios are not given, and must be chosen; the eavesdropping ratios $\bmu=(\mu_1, \cdots, \mu_N)$ are given here. Our converse proof extends the proof in \cite{KarimAsymmetricPIR} to account for the imposed security constraint. 

In the next lemma, we discuss some consequences of the security constraint in \eqref{security_constraint}. The security constraint introduces some interesting conditional independence properties which simplify the converse proof.

\begin{lemma}[Security consequences]\label{lemma 0}
	In the PIR-WTC-II problem, the following implications are true due to the security constraint \eqref{security_constraint}:
	
	\begin{enumerate}
		\item Messages are conditionally independent given the observed part of the answer strings at the eavesdropper $Z_{1:N}^{[i]}$, i.e.,
		\begin{align}\label{lemma0_1}
		I(W_m;W_{[1:M]\setminus\{m\}}|Z_{1:N}^{[i]})=0, \quad i,\, m \in \{1, \cdots, M\}
		\end{align}
		\item There is no leakage of $W_m$ from all the queries $Q_{1:N}^{[i]}$, the eavesdropper observations $Z_{1:N}^{[i]}$, and any subset of messages $W_\mathcal{S}=\{W_i: i \in \mathcal{S}\}$ such that $m \notin \mathcal{S}$,
		\begin{align}\label{lemma0_2}
		I(W_m;W_\mathcal{S},Z_{1:N}^{[i]},Q_{1:N}^{[i]})=0, \quad i,\,m \in \{1, \cdots, M\}
		\end{align}
		In particular, 
		\begin{align}\label{lemma0_22}
		I(W_m;W_{m:M}|W_{1:m-1},Z_{1:N}^{[i]})=L, \quad i,\,m \in \{1, \cdots, M\}
		\end{align}
		\item The eavesdropper's observations $Z_{1:N}^{[i]}$ and the messages are conditionally independent given the queries $Q_{1:N}^{[i]}$, i.e., for sets $\mathcal{S}_1$, $\mathcal{S}_2$, such that $\mathcal{S}_1 \cap \mathcal{S}_2=\emptyset$,
		\begin{align}\label{lemma0_3}
		I(W_{\mathcal{S}_1};Z_{1:N}^{[i]}|Q_{1:N}^{[i]},W_{\mathcal{S}_2})=0, \quad i \in \{1,\cdots,M\}
		\end{align}
		In particular,
		\begin{align}\label{lemma0_31}
		I(W_{m:M};Z_{1:N}^{[m-1]}|W_{1:m-1})=0, \quad m \in \{2,\cdots,M\}
		\end{align}
		\item The messages and the queries are conditionally independent given the eavesdropper's observations, i.e., for sets $\mathcal{S}_1$, $\mathcal{S}_2$, such that $\mathcal{S}_1 \cap \mathcal{S}_2=\emptyset$,
		\begin{align}\label{lemma0_4}
		I(W_{\mathcal{S}_1};Q_{1:N}^{[i]}|W_{\mathcal{S}_2},Z_{1:N}^{[i]})=0,\quad i \in \{1,\cdots,M\}
		\end{align}
		\item The messages $W_{m:M}$ and the queries $Q_{k+1:N}^{[m]}$ for any $k \in \{1, \cdots, N\}$ are conditionally independent given $\left(W_{1:m-1},Z_{1:N}^{[m]},Q_{1:k}^{[m]},Y_{1:k}^{[m]}\right)$, i.e.,
		\begin{align}\label{lemma0_5}
		I(W_{m:M};Q_{k+1:N}^{[m]}|W_{1:m-1},Z_{1:N}^{[m]},Q_{1:k}^{[m]},Y_{1:k}^{[m]})=0
		\end{align}
	\end{enumerate}
\end{lemma}

\begin{Proof}
	\begin{enumerate}
		\item From the security constraint \eqref{security_constraint}, we have $I(W_{1:M};Z_{1:N}^{[i]},Q_{1:N}^{[i]})=0$, which further implies that $I(W_{1:M};Z_{1:N}^{[i]})=0$. This can be expanded as:
		\begin{align}
		0&=I(W_m,W_{[1:M]\setminus\{m\}};Z_{1:N}^{[i]}) \label{l0_1}\\
		&=I(W_m;Z_{1:N}^{[i]})+I(W_{[1:M]\setminus\{m\}};Z_{1:N}^{[i]}|W_m) \label{l0_2}\\
		&=I(W_{[1:M]\setminus\{m\}};Z_{1:N}^{[i]})+I(W_{m};Z_{1:N}^{[i]}|W_{[1:M]\setminus\{m\}}) \label{l0_3}
		\end{align}
		which implies that all four terms in \eqref{l0_2}, \eqref{l0_3} are zero. Then, consider
		\begin{align}
		I(W_m;W_{[1:M]\setminus\{m\}},Z_{1:n}^{[i]})&=I(W_m;Z_{1:N}^{[i]})+I(W_m;W_{[1:M]\setminus\{m\}}|Z_{1:N}^{[i]}) \\
		&=I(W_m;W_{[1:M]\setminus\{m\}})+I(W_m;Z_{1:N}^{[i]}|W_{[1:M]\setminus \{m\}})
		\end{align}
		which together with \eqref{l0_2}, \eqref{l0_3} and the independence of the messages imply \eqref{lemma0_1}. 
		
		\item From the security constraint \eqref{security_constraint}, we have $I(W_m,W_\mathcal{S};Q_{1:N}^{[i]},Z_{1:N}^{[i]})=0$ by the non-negativity of mutual information. This can be further expanded as
		\begin{align}
		0=I(W_m,W_\mathcal{S};Q_{1:N}^{[i]},Z_{1:N}^{[i]})=I(W_\mathcal{S};Q_{1:N}^{[i]},Z_{1:N}^{[i]})+I(W_m;Q_{1:N}^{[i]},Z_{1:N}^{[i]}|W_\mathcal{S})
		\end{align}
		From the second term on the right hand side, we have $I(W_m;Q_{1:N}^{[i]},Z_{1:N}^{[i]}|W_\mathcal{S})=0$, which implies \eqref{lemma0_2} by the independence of the messages, as $I(W_m;W_\mathcal{S},Z_{1:N}^{[i]},Q_{1:N}^{[i]})=I(W_m; W_\mathcal{S})+I(W_m;Z_{1:N}^{[i]},Q_{1:N}^{[i]}|W_\mathcal{S})$.
		
		For \eqref{lemma0_22}, we note that \eqref{lemma0_2} implies that $I(W_m;W_{1:m-1},Z_{1:N}^{[i]})=0$ by the non-negativity of mutual information, which further implies that $I(W_m;Z_{1:N}^{[i]}|W_{1:m-1})=0$. Now,
		\begin{align}
		I(W_m;W_{m:M}|W_{1:m-1},Z_{1:N}^{[i]})=&H(W_m|W_{1:m-1},Z_{1:N}^{[i]})\\
		=&H(W_m|W_{1:m-1})-I(W_m;Z_{1:N}^{[i]}|W_{1:m-1})\\
		=&L
		\end{align} 
		where the last equality follows from the independence of the messages.
		
		\item From the security constraint \eqref{security_constraint} and the non-negativity of mutual information, we have $I(W_{\mathcal{S}_1},W_{\mathcal{S}_2};Z_{1:N}^{[i]},Q_{1:N}^{[i]})=0$, which can be expanded as $I(W_{\mathcal{S}_2};Z_{1:N}^{[i]},Q_{1:N}^{[i]})+I(W_{\mathcal{S}_1};Z_{1:N}^{[i]},Q_{1:N}^{[i]}|W_{\mathcal{S}_2})=0$, which implies that $I(W_{\mathcal{S}_1};Z_{1:N}^{[i]},Q_{1:N}^{[i]}|W_{\mathcal{S}_2})=0$. We futher expand it as:
		\begin{align}\label{l0_4}
		0=I(W_{\mathcal{S}_1};Q_{1:N}^{[i]}|W_{\mathcal{S}_2})+I(W_{\mathcal{S}_1};Z_{1:N}^{[i]}|Q_{1:N}^{[i]},W_{\mathcal{S}_2})
		\end{align}
		which leads to \eqref{lemma0_3} by the non-negativity of mutual information.
		
		For \eqref{lemma0_31}, we note from \eqref{lemma0_3} that $I(W_{m:M};Z_{1:N}^{[m-1]}|Q_{1:N}^{[m-1]},W_{1:m-1})=0$, hence
		\begin{align}
		0=I(W_{m:M};Z_{1:N}^{[m-1]},Q_{1:N}^{[m-1]}|W_{1:m-1})-I(W_{m:M};Q_{1:N}^{[m-1]}|W_{1:m-1})
		\end{align}
		Now, $I(W_{m:M};Q_{1:N}^{[m-1]}|W_{1:m-1})=0$ by the independence of the messages and the queries in \eqref{independency}, and this implies \eqref{lemma0_31} by the non-negativity of mutual information.
		\item Using the same argument as in item 3 above and reversing the order of the chain rule in \eqref{l0_4} leads to \eqref{lemma0_4}.
		\item We have
		\begin{align}
		I(&W_{m:M};Q_{k+1:N}^{[m]}|W_{1:m-1},Z_{1:N}^{[m]},Q_{1:k}^{[m]},Y_{1:k}^{[m]})\notag\\
		=&I(W_{m:M};Q_{k+1:N}^{[m]},Y_{1:k}^{[m]}|W_{1:m-1},Z_{1:N}^{[m]},Q_{1:k}^{[m]})-I(W_{m:M};Y_{1:k}^{[m]}|W_{1:m-1},Z_{1:N}^{[m]},Q_{1:k}^{[m]})\\
		=&I(W_{m:M};Q_{k+1:N}^{[m]}|W_{1:m-1},Z_{1:N}^{[m]},Q_{1:k}^{[m]})+I(W_{m:M};Y_{1:k}^{[m]}|W_{1:m-1},Z_{1:N}^{[m]},Q_{1:N}^{[m]})\notag\\
		&-I(W_{m:M};Y_{1:k}^{[m]}|W_{1:m-1},Z_{1:N}^{[m]},Q_{1:k}^{[m]})\\
		=&0
		\end{align}
		where $I(W_{m:M};Q_{k+1:N}^{[m]}|W_{1:m-1},Z_{1:N}^{[m]},Q_{1:k}^{[m]})=0$ from \eqref{lemma0_4} and the non-negativity of mutual information, and since $Q_{1:N}^{[m]} \rightarrow Q_{1:k}^{[m]} \rightarrow Y_{1:k}^{[m]}$ is a Markov chain, we have
		$I(W_{m:M};Y_{1:k}^{[m]}|W_{1:m-1},Z_{1:N}^{[m]},Q_{1:N}^{[m]})=I(W_{m:M};\!Y_{1:k}^{[m]}|W_{1:m-1},Z_{1:N}^{[m]},Q_{1:k}^{[m]})$.
	\end{enumerate}
\end{Proof}

We will need the following lemma, which characterizes a lower bound on the interference from the undesired messages within the portion of answers that is unobserved by the eavesdropper (and hence secure). Since the user must download at least $L$ symbols to retrieve the desired message, the difference $\sum_{n=1}^N (1-\mu_n)t_n-L$  denotes the interference terms within the unobserved (by the eavesdropper) portion of the answers.

\begin{lemma}[Interference lower bound]\label{lemma_converse1}
	For the PIR-WTC-II problem, the interference from undesired messages within the unobserved portion of the answer strings by the eavesdropper  $\sum_{n=1}^N (1-\mu_n)t_n-L$ is lower bounded by, 	
	\begin{align}
	\sum_{n=1}^N (1-\mu_n)t_n-L+ o(L) \geq I\left(W_{2:M};Q_{1:N}^{[1]}, Y_{1:N}^{[1]}|W_{1}, Z_{1:N}^{[1]} \right) \label{eq_L1}
	\end{align}
\end{lemma}

We note that Lemma~\ref{lemma_converse1} is a generalization of \cite[Lemma~5]{JafarPIR} to the problem of PIR-WTC-II. If $\mu_n=0$ for all $n \in [1:N]$, then Lemma~\ref{lemma_converse1} reduces to \cite[Lemma~5]{JafarPIR} as $Z_{1:N}^{[1]}$ (the eavesdropper observations) is absent and $Y_{1:N}^{[1]}=A_{1:N}^{[i]}$ in that case.

\begin{Proof}
	We start with the right hand side of \eqref{eq_L1},
	\begin{align}
		I(W_{2:M}&;Q_{1:N}^{[1]},Y_{1:N}^{[1]}|W_1,Z_{1:N}^{[1]})\notag\\
		\stackrel{\eqref{lemma0_1}}{=}& I\left(W_{2:M};W_1,Q_{1:N}^{[1]},Y_{1:N}^{[1]}|Z_{1:N}^{[1]}\right) \label{L1_1} \\
	    =&I\left(W_{2:M};Q_{1:N}^{[1]},Y_{1:N}^{[1]}|Z_{1:N}^{[1]}\right)+I\left(W_{2:M};W_1|A_{1:N}^{[1]},Q_{1:N}^{[1]}\right)  \\
	    \stackrel{\eqref{reliability_constraint}}{=}& I\left(W_{2:M};Q_{1:N}^{[1]},Y_{1:N}^{[1]}|Z_{1:N}^{[1]}\right)+o(L) \label{L1_2}\\ 
	    \stackrel{\eqref{lemma0_4}}{=}& I\left(W_{2:M};Y_{1:N}^{[1]}|Q_{1:N}^{[1]},Z_{1:N}^{[1]}\right)+o(L) \label{L1_3}\\  
	    =&H\left(Y_{1:N}^{[1]}|Q_{1:N}^{[1]},Z_{1:N}^{[1]}\right)-H\left(Y_{1:N}^{[1]}|Q_{1:N}^{[1]},Z_{1:N}^{[1]},W_{2:M}\right)+o(L) \\
	    \leq &\!\sum_{n=1}^N (1-\mu_n)t_n\!-\!H\left(\!W_1,Y_{1:N}^{[1]}|Q_{1:N}^{[1]},\!Z_{1:N}^{[1]},\!W_{2:M}\!\right)\!+\!H\left(\!W_1|A_{1:N}^{[1]},\!Q_{1:N}^{[1]},\!W_{2:M}\!\right)\!+\!o(L) \label{L1_4}\\
	    \stackrel{\eqref{reliability_constraint}}{=}&\sum_{n=1}^N (1-\mu_n)t_n-H\left(W_1,Y_{1:N}^{[1]}|Q_{1:N}^{[1]},Z_{1:N}^{[1]},W_{2:M}\right)+o(L) \label{L1_5}\\
	    =&\sum_{n=1}^N (1-\mu_n)t_n\!-\!H\left(W_1|Q_{1:N}^{[1]},Z_{1:N}^{[1]},W_{2:M}\right)\!-\!H\left(Y_{1:N}^{[1]}|Q_{1:N}^{[1]},Z_{1:N}^{[1]},W_{1:M}\right)\!+\!o(L)\\
	    \leq &\sum_{n=1}^N (1-\mu_n)t_n\!-\!H\left(W_1|Q_{1:N}^{[1]},Z_{1:N}^{[1]}, W_{2:M}\right)+o(L) \label{L1_6}\\
	    \stackrel{\eqref{lemma0_2}}{=}&\sum_{n=1}^N (1-\mu_n)t_n-L+o(L) \label{L1_7} 
	\end{align}
	where \eqref{L1_1} follows from the conditional independence of messages in Lemma~\ref{lemma 0}, \eqref{L1_2},\,\eqref{L1_5} follow from the decodability of $W_1$ given $(Q_{1:N}^{[1]},A_{1:N}^{[1]})$, \eqref{L1_3} follows from the conditional independence of the messages and the queries in Lemma~\ref{lemma 0}, \eqref{L1_4} follows from conditioning reduces entropy and the fact that $H(Y_{1:N}^{[1]}) \leq \sum_{n=1}^{N} (1-\mu_n)t_n$ from the WTC-II model, \eqref{L1_6} follows from the non-negativity of the entropy function, and \eqref{L1_7} follows from zero leakage property of $W_1$ from \eqref{lemma0_2} which implies $H(W_1|Q_{1:N}^{[1]},Z_{1:N}^{[1]},W_{2:M})=H(W_1)=L$.  
\end{Proof}


In the following lemma, we derive an induction relation for the right hand side of the expression in \eqref{eq_L1}. This lemma extends \cite[Lemma~6]{JafarPIR} in two major ways. First, we incorporate the security constraint in the proof by observing that $(W_{1:M},Z_{1:N}^{[m]})$ are independent. Second, and more significantly, the main difference between this lemma and \cite[Lemma~6]{JafarPIR} is the fact that not all databases can use a symmetric scheme due to the asymmetry of the fraction that the eavesdropper can observe. Consequently, we denote $n_{m-1}$ to be the number of databases that can apply a symmetric scheme when the retrieval problem is reduced to retrieving message $W_{m-1}$ from the set of $W_{m-1:M}$ messages. For the remaining answer strings, we directly bound them by their corresponding length of the unobserved portion $\sum_{n=n_{m-1}+1}^N (1-\mu_n)t_n$. 

\begin{lemma}[Induction lemma]\label{lemma_converse2}
	For all $m\in \{2,\dots,M\}$ and for an arbitrary $n_{m-1} \in \{1, \cdots, N\}$, the mutual information term in Lemma~\ref{lemma_converse1} can be inductively lower bounded as,
	\begin{align} \label{eq_L2}
	&I\left( W_{m:M} ; Q_{1:N}^{[m-1]}, Y_{1:N}^{[m-1]} | W_{1:m-1},Z_{1:N}^{[m-1]} \right)  \notag \\
	&\quad\geq \frac{1}{n_{m-1}}\left[I\left(W_{m+1:M}; Q_{1:N}^{[m]}, Y_{1:N}^{[m]}|W_{1:m},Z_{1:N}^{[m]} \right)  + \left(L-\!\!\!\!\!\sum_{n=n_{m-1}+1}^N\!\!\! (1-\mu_n)t_n \right)-o(L)\right]
	\end{align}
\end{lemma}

\begin{Proof}  
	We start with the left hand side of \eqref{eq_L2}, after multiplying by $n_{m-1}$,
	\begin{align}
	& n_{m-1}\,I\left(W_{m:M} ; Q_{1:N}^{[m-1]}, Y_{1:N}^{[m-1]}| W_{1:m-1},Z_{1:N}^{[m-1]}  \right) \notag \\
	&\label{eq_IL_11}\quad \stackrel{\eqref{lemma0_31}}{=} n_{m-1}\,I\left(W_{m:M} ; Q_{1:N}^{[m-1]}, A_{1:N}^{[m-1]}| W_{1:m-1}  \right)\\
	&\quad \geq  n_{m-1}\, I\left(W_{m:M} ; Q_{1:n_{m-1}}^{[m-1]}, A_{1:n_{m-1}}^{[m-1]}| W_{1:m-1}  \right) \label{eq_IL_12}\\
	& \label{eq_IL_1}\quad \geq  \sum_{n=1}^{n_{m-1}} I\left(W_{m:M} ; Q_n^{[m-1]}, A_n^{[m-1]}| W_{1:m-1}  \right) \\
	& \label{eq_IL_2}\quad \stackrel{\eqref{privacy_constraint}}{=}  \sum_{n=1}^{n_{m-1}} I\left(W_{m:M} ; Q_n^{[m]}, A_n^{[m]}| W_{1:m-1}  \right) \\ 
	& \label{eq_IL_444}\quad \stackrel{\eqref{independency}}{=}  \sum_{n=1}^{n_{m-1}} I\left(W_{m:M} ;  A_n^{[m]}|Q_n^{[m]}, W_{1:m-1}  \right)\\
	& \label{eq_IL_45}\quad \stackrel{\eqref{lemma0_3}}{=}  \sum_{n=1}^{n_{m-1}} I\left(W_{m:M} ;  Y_n^{[m]}|Q_n^{[m]}, W_{1:m-1}, Z_n^{[m]}  \right)\\
	& \label{eq_IL_44}\quad =  \sum_{n=1}^{n_{m-1}} H\left(Y_n^{[m]}|Q_n^{[m]}, W_{1:m-1},Z_n^{[m]}  \right)-H\left(Y_n^{[m]}|Q_n^{[m]}, W_{1:M},Z_n^{[m]}  \right) \\
	& \label{eq_IL_3}\quad \geq \sum_{n=1}^{n_{m-1}}\!\! H\left(\!Y_n^{[m]}|Y^{[m]}_{1:n-1},Q_{1:n_{m-1}}^{[m]}, W_{1:m-1},Z_{1:N}^{[m]}  \!\right)\!-\!H\left(Y_n^{[m]}|Y^{[m]}_{1:n-1},Q_{1:n_{m-1}}^{[m]}, W_{1:M},Z_{1:N}^{[m]}\right) \\ 
	& \label{eq_IL_4}\quad = \sum_{n=1}^{n_{m-1}} I\left(W_{m:M};Y_n^{[m]}|Y^{[m]}_{1:n-1},Q_{1:n_{m-1}}^{[m]}, W_{1:m-1},Z_{1:N}^{[m]}   \right) \\
	&\label{eq_IL_5}  \quad = I\left(W_{m:M}; Y_{1:n_{m-1}}^{[m]} | Q_{1:n_{m-1}}^{[m]}, W_{1:m-1},Z_{1:N}^{[m]}  \right) \\
	&\label{eq_IL_6} \quad \stackrel{\eqref{lemma0_4}}{=} I\left(W_{m:M}; Q_{1:n_{m-1}}^{[m]}, Y_{1:n_{m-1}}^{[m]} |  W_{1:m-1},Z_{1:N}^{[m]}  \right)\\
	&\label{eq_IL_65}\quad \stackrel{\eqref{lemma0_5}}{=}I\left(W_{m:M}; Q_{1:N}^{[m]}, Y_{1:N}^{[m]} |  W_{1:m-1},Z_{1:N}^{[m]}\right)\notag\\
	&\quad\quad\:-I\left(W_{m:M}; Y_{n_{m-1}+1:N}^{[m]} | Q_{1:N}^{[m]}, Y_{1:n_{m-1}}^{[m]}, W_{1:m-1},Z_{1:N}^{[m]}  \right)\\
	&\quad \label{eq_IL_66} \geq I\!\left(\!W_{m:M}; Q_{1:N}^{[m]}, Y_{1:N}^{[m]} |  W_{1:m-1},Z_{1:N}^{[m]}\right)\!-\!H\left(Y_{n_{m-1}+1:N}^{[m]} \right)\\
	&\quad \label{eq_IL_67}\geq I\left(W_{m:M}; Q_{1:N}^{[m]}, Y_{1:N}^{[m]} |  W_{1:m-1},Z_{1:N}^{[m]}\right)-\!\!\!\sum_{n=n_{m-1}+1}^{N}\!\!\! (1-\mu_n)t_n\\
	& \quad= I\left(W_{m:M}; W_m, Q_{1:N}^{[m]}, Y_{1:N}^{[m]} |  W_{1:m-1},Z_{1:N}^{[m]}\right)-I\left(W_{m:M}; W_m|  W_{1:m-1},Q_{1:N}^{[m]},A_{1:N}^{[m]}\right)\notag\\
	&\quad \quad -\!\!\!\sum_{n=n_{m-1}+1}^{N}\!\!\! (1-\mu_n)t_n\\
	& \label{eq_IL_7} \quad \stackrel{\eqref{reliability_constraint}}{=} I\left(W_{m:M}; W_m, Q_{1:N}^{[m]}, Y_{1:N}^{[m]} |  W_{1:m-1},Z_{1:N}^{[m]}\right)-\!\!\!\sum_{n=n_{m-1}+1}^{N}\!\!\! (1-\mu_n)t_n-o(L)\\
	& \quad =I\left(W_{m:M};W_m|  W_{1:m-1},Z_{1:N}^{[m]}\right)+I\left(W_{m:M};Q_{1:N}^{[m]}, Y_{1:N}^{[m]} |  W_{1:m},Z_{1:N}^{[m]}\right)\notag\\
	&\quad\quad-\!\!\!\sum_{n=n_{m-1}+1}^{N}\!\!\! (1-\mu_n)t_n -o(L)\\
	& \label{L2_final}\quad \stackrel{\eqref{lemma0_22}}{=}I\left(W_{m+1:M}; Q_{1:N}^{[m]}, Y_{1:N}^{[m]}|W_{1:m},Z_{1:N}^{[m]} \right)  + \left(L-\!\!\!\sum_{n=n_{m-1}+1}^{N}\!\!\! (1-\mu_n)t_n\right)-o(L)
	\end{align}
	where \eqref{eq_IL_11} follows from the conditional independence of the messages and $Z_{1:N}^{[m-1]}$ in \eqref{lemma0_31} as a consequence of the security constraint, \eqref{eq_IL_12}, \eqref{eq_IL_1} follow from the non-negativity of mutual information, \eqref{eq_IL_2} follows from the privacy constraint, \eqref{eq_IL_444} follows from the independence of the queries and the messages,  \eqref{eq_IL_45} follows from the conditional independence of the messages and $Z_n^{[m]}$ in \eqref{lemma0_3} and the non-negativity of mutual information, \eqref{eq_IL_3} follows from conditioning reduces entropy and $\left(Q_{1:n_{m-1}}^{[m]},Z_{1:N}^{[m]},W_{1:M},Y_{1:n-1}^{[m]}\right) \rightarrow \left(Q_n^{[m]},W_{1:M},Z_n^{[n]}\right) \rightarrow Y_n^{[m]}$, \eqref{eq_IL_6} follows from \eqref{lemma0_4} and the non-negativity of mutual information, \eqref{eq_IL_65} follows from the chain rule and \eqref{lemma0_5}, \eqref{eq_IL_66} follows from the fact that $I\left(W_{m:M}; Y_{n_{m-1}+1:N}^{[m]} | Q_{1:N}^{[m]}, Y_{1:n_{m-1}}^{[m]}, W_{1:m-1},Z_{1:N}^{[m]}\right) \leq H\left(Y_{1:n_{m-1}}^{[m]}\right)$, \eqref{eq_IL_67} follows from the fact that conditioning reduces entropy and $H(Y_{n_{m-1}+1:N}^{[m]})\leq \sum_{n=n_{m-1}+1}^N (1-\mu_n)t_n$ in the WTC-II model, \eqref{eq_IL_7} follows from the reliability constraint, \eqref{L2_final} follows from the no leakage property of $W_m$ from \eqref{lemma0_22} as a consequence of the security constraint. Finally, dividing both sides by $n_{m-1}$ leads to \eqref{eq_L2}.
\end{Proof}

Now, we are ready to prove an explicit upper bound for the retrieval rate in the PIR-WTC-II problem $R(\boldsymbol{\mu})$ by applying Lemma~\ref{lemma_converse1} and Lemma~\ref{lemma_converse2} successively. For a pre-specified answer string lengths $\{t_n\}_{n=1}^N$, and an arbitrary sequence $\{n_i\}_{i=1}^{M-1}$, we can write
\begin{align}
&\sum_{n=1}^N (1-\mu_n)t_n-L+\tilde{o}(L)  \notag \\
&\label{eq_induction1}\quad \stackrel{\eqref{eq_L1}}{\geq} I\left(W_{2:M}; Q_{1:N}^{[1]}, Y_{1:N}^{[1]}|W_{1},Z_{1:N}^{[1]} \right) \\
&\quad \stackrel{\eqref{eq_L2}}{\geq}\frac{1}{n_{1}} \left(L-\!\!\!\!\!\sum_{n=n_{1}+1}^N\!\!\! (1-\mu_n)t_n \right)+ \frac{1}{n_{1}}  I\left(W_{3:M}; Q_{1:N}^{[2]}, Y_{1:N}^{[2]}|W_{1:2},Z_{1:N}^{[2]} \right)    \\
&\quad \stackrel{\eqref{eq_L2}}{\geq}\frac{1}{n_{1}} \left(L-\!\!\!\!\!\sum_{n=n_{1}+1}^N\!\!\! (1-\mu_n)t_n \right)+\frac{1}{n_{1}n_2} \left(L-\!\!\!\!\!\sum_{n=n_{2}+1}^N\!\!\! (1-\mu_n)t_n \right)\notag\\
&\quad\quad\:\:+ \frac{1}{n_{2}}  I\left(W_{4:M}; Q_{1:N}^{[3]}, Y_{1:N}^{[3]}|W_{1:3},Z_{1:N}^{[3]} \right)    \\
&\quad \stackrel{\eqref{eq_L2}}{\geq} \dots \notag\\
&\quad \stackrel{\eqref{eq_L2}}{\geq}\frac{1}{n_{1}} \left(L-\!\!\!\!\!\sum_{n=n_{1}+1}^N\!\!\! (1-\mu_n)t_n \right)+\frac{1}{n_{1}n_2} \left(L-\!\!\!\!\!\sum_{n=n_{2}+1}^N\!\!\! (1-\mu_n)t_n \right)+ \cdots \notag\\
&\quad\quad \!+\!\frac{1}{\prod_{i=1}^{M-1} n_i}\!\left(L-\!\!\!\!\!\sum_{n=n_{M-1}+1}^N\!\!\! (1-\mu_n)t_n \right)
\end{align}
where $\tilde{o}(L)=\left(1+\frac{1}{n_1}+\frac{1}{n_1n_2}+\cdots+\frac{1}{\prod_{i=1}^{M-1}n_i}\right)o(L)$, \eqref{eq_induction1} follows from Lemma~\ref{lemma_converse1}, and the remaining bounding steps follow from successive application of Lemma~\ref{lemma_converse2}.

Ordering terms and letting $\tau_n=\frac{t_n}{\sum_{i=1}^{N} t_i}$, we have, 
\begin{align}
\left(1+\frac{1}{n_1}\!+\!\frac{1}{n_1n_2}\!+\!\cdots\!+\!\frac{1}{\prod_{i=1}^{M-1}n_i}\right)\!L \leq 
\left(\phi(0)+\frac{\phi(n_1)}{n_1}+\!\cdots\!+\frac{\phi(n_{M-1})}{\prod_{i=1}^{M-1} n_i}\right)\!\sum_{n=1}^N t_n \!+\! \tilde{o}(L)
\end{align}
where $\phi(\ell)=\sum_{n=\ell+1}^N (1-\mu_n)\tau_n$ corresponds to the sum of the unobserved traffic ratios by the eavesdropper from databases $[\ell+1:N]$.

We conclude the proof by taking $L \rightarrow \infty$. Thus, for an arbitrary sequence $\{n_i\}_{i=1}^{M-1}$ we have
\begin{align}\label{final_converse}
R(\boldsymbol{\tau},\boldsymbol{\mu})&= \frac{L}{\sum_{n=1}^{N} t_n} \leq  \frac{\phi(0)+\frac{\phi(n_1)}{n_1}+\frac{\phi(n_2)}{n_1n_2}+\cdots+\frac{\phi(n_{M-1})}{\prod_{i=1}^{M-1} n_i}}{1+\frac{1}{n_1}+\frac{1}{n_1n_2}+\cdots+\frac{1}{\prod_{i=1}^{M-1}n_i}}
\end{align}
The bound in \eqref{final_converse} for $R(\boldsymbol{\tau},\boldsymbol{\mu})$ is valid for any arbitrary sequence $\{n_i\}_{i=1}^{M-1}$. Hence, we obtain the tightest upper bound for $R(\boldsymbol{\tau},\boldsymbol{\mu})$ by minimizing over the sequence $\{n_i\}_{i=1}^{M-1}$ over the set $\{1, \cdots, N\}$ to get

\begin{align}
R(\boldsymbol{\tau},\boldsymbol{\mu}) &\leq \min_{n_1, \cdots, n_{M-1} \in \{1, \cdots, N\}} \frac{\phi(0)+\frac{\phi(n_1)}{n_1}+\frac{\phi(n_2)}{n_1n_2}+\cdots+\frac{\phi(n_{M-1})}{\prod_{i=1}^{M-1} n_i}}{1+\frac{1}{n_1}+\frac{1}{n_1n_2}+\cdots+\frac{1}{\prod_{i=1}^{M-1}n_i}}
\end{align}

Finally, since the user and the databases can choose any suitable traffic ratio vector $\boldsymbol{\tau}$ in the set $\mathbb{T}$ such that:
\begin{align}
\mathbb{T}=\left\{\boldsymbol{\tau}: \tau_n \geq 0 \quad \forall n \in [1:N],\quad \sum_{n=1}^{N} \tau_n=1\right\}
\end{align}
by maximizing over $\boldsymbol{\tau}=(\tau_1,\tau_2, \cdots, \tau_N)$ in the set $\mathbb{T}$, we obtain the following upper bound for $R(\boldsymbol{\mu})$,

\begin{align}
R(\boldsymbol{\mu}) &\leq \max_{\boldsymbol{\tau} \in \mathbb{T}} \min_{n_i \in \{1, \cdots, N\}} \frac{\phi(0)+\frac{\phi(n_1)}{n_1}+\frac{\phi(n_2)}{n_1n_2}+\cdots+\frac{\phi(n_{M-1})}{\prod_{i=1}^{M-1} n_i}}{1+\frac{1}{n_1}+\frac{1}{n_1n_2}+\cdots+\frac{1}{\prod_{i=1}^{M-1}n_i}}\\
&=\max_{\boldsymbol{\tau} \in \mathbb{T}} \min_{n_i \in \{1, \cdots, N\}} \frac{\sum_{n=1}^N (1-\mu_n)\tau_n+\frac{\sum_{n=n_1+1}^N (1-\mu_n)\tau_n}{n_1}+\cdots+\frac{\sum_{n=n_{M-1}+1}^N (1-\mu_n)\tau_n}{\prod_{i=1}^{M-1} n_i}}{1+\frac{1}{n_1}+\cdots+\frac{1}{\prod_{i=1}^{M-1}n_i}}
\end{align}

\section{Achievable Scheme}\label{achievability}
In this section, we present a general achievable scheme for PIR-WTC-II. The scheme builds on the achievable scheme in \cite{KarimAsymmetricPIR}. The main idea of the achievable scheme is that since the databases are eavesdropped by varying eavesdropping capabilities $\bmu$, then it would be beneficial for the user to query the databases using the PIR scheme under asymmetric traffic constraints. Furthermore, the databases should \emph{encrypt} the answers such that the user can decode the \emph{meaningful} transmission by observing the entire answer string, while the encryption keys span the eavesdropper's entire observation space, ensuring the security of downloaded content. The user and the databases agree on the traffic ratio vector $\bt$ that maximizes the achievable secure PIR rate.

In the following, we illustrate the main ingredients of the achievable scheme by presenting the case of $M=3$ messages and $N=2$ databases for an arbitrary $\bmu$.  

\subsection{Motivating Example: $M=3$ Messages, $N=2$ Databases}
In this section, we first show an explicit upper bound for the capacity expression $\bar{C}(\bmu)$. Then, we show the capacity-achieving scheme for the concrete example of $\bmu=(\frac{1}{4},\frac{1}{2})$. We conclude this section by showing how to extend the achievable scheme for arbitrary $\bmu$.
\subsubsection{Explicit Upper Bound for $M=3$ Messages, $N=2$ Databases} 
From Theorem \ref{Thm1}, the upper bound of $\bar{C}(\bmu)$ is given by:
\begin{align}
\bar{C}(\bmu)=\max_{\boldsymbol{\tau} \in \mathbb{T}} \min_{n_i \in \{1,2\}} \frac{\sum_{n=1}^2 (1-\mu_n)\tau_n+\frac{\sum_{n=n_1+1}^2 (1-\mu_n)\tau_n}{n_1}+\frac{\sum_{n=n_2+1}^2 (1-\mu_n)\tau_n}{n_1n_2 }}{1+\frac{1}{n_1}+\frac{1}{n_1n_2}}
\end{align}

By observing that $\tau_1=1-\tau_2$, this can be explicitly written as the following linear program:
    \begin{align}\label{LP32}
    \max_{\tau_2,R}  &\quad R \notag\\
    \st &\quad R \leq \frac{1}{3}(1-\mu_1)+\left[(1-\mu_2)-\frac{1}{3}(1-\mu_1)\right]\tau_2\notag\\
        &\quad R \leq \frac{2}{5}(1-\mu_1)+\left[\frac{4}{5}(1-\mu_2)-\frac{2}{5}(1-\mu_1)\right]\tau_2\notag\\
        &\quad R \leq \frac{4}{7}(1-\mu_1)+\left[\frac{4}{7}(1-\mu_2)-\frac{4}{7}(1-\mu_1)\right]\tau_2\notag\\
        &\quad 0 \leq \tau_2 \leq 1
    \end{align}
Note that the bound corresponding to $n_1=2,n_2=1$ is not included in \eqref{LP32} as it would be inactive for any $\bmu$. Since \eqref{LP32} is a linear program, the optimal solution exists among the corner points of the feasible region. The first corner point, is $\tau_2^{(1)}=0$, which leads to the bound $\bar{C}(\bmu) \leq \frac{1-\mu_1}{3}$. The second corner point occurs at the intersection of the first two constraints, i.e., $\tau_2^{(2)}$ satisfies:
\begin{align}
\frac{1}{3}(1-\mu_1)+\left[(1-\mu_2)-\frac{1}{3}(1-\mu_1)\right]\tau_2^{(2)}=\frac{2}{5}(1-\mu_1)+\left[\frac{4}{5}(1-\mu_2)-\frac{2}{5}(1-\mu_1)\right]\tau_2^{(2)}
\end{align}
which leads to,
\begin{align}
\tau_2^{(2)}=\frac{(1-\mu_1)}{3(1-\mu_2)+(1-\mu_1)}
\end{align}
with a corresponding bound of $\bar{C}(\bmu) \leq \frac{2(1-\mu_1)(1-\mu_2)}{3(1-\mu_2)+(1-\mu_1)}$. Similarly, the third corner point $\tau_2^{(3)}$ occurs at the intersection of the second and third constraints, hence $\tau_2^{(3)}=\frac{3(1-\mu_1)}{4(1-\mu_2)+3(1-\mu_1)}$ with the corresponding bound of $\bar{C}(\bmu) \leq \frac{4(1-\mu_1)(1-\mu_2)}{4(1-\mu_2)+3(1-\mu_1)}$. Finally, at $\tau_2=1$, we have the bound $\bar{C}(\bmu) \leq \frac{4(1-\mu_2)}{7}$ which is no larger than $\frac{4(1-\mu_1)(1-\mu_2)}{4(1-\mu_2)+3(1-\mu_1)}$ by the monotonicity of $\bmu$, hence it can be ignored.  

Consequently, the explicit upper bound for $M=3$, $N=2$ is given by
\begin{align}\label{ub32}
\bar{C}(\bmu)=\max \left\{\frac{1-\mu_1}{3},\:\frac{2(1-\mu_1)(1-\mu_2)}{3(1-\mu_2)+(1-\mu_1)},\: \frac{4(1-\mu_1)(1-\mu_2)}{4(1-\mu_2)+3(1-\mu_1)} \right\}
\end{align}
\subsubsection{Concrete Example: $\mu_1=\frac{1}{4}$, $\mu_2=\frac{1}{2}$}\label{example}
Before the retrieval process, the user permutes the indices of the symbols of $W_1$, $W_2$, $W_3$ independently, uniformly, and privately. Assume without loss of generality that $W_1$ is the desired message. Let $a_i$, $b_i$, $c_i$ be the permuted symbols from $W_1$, $W_2$, $W_3$, respectively. In the case of $\mu_1=\frac{1}{4}$, $\mu_2=\frac{1}{2}$, the explicit upper bound in \eqref{ub32} is $\bar{C}(\bmu)=\frac{4(1-\mu_1)(1-\mu_2)}{4(1-\mu_2)+3(1-\mu_1)}=\frac{6}{17}$. To achieve this bound, we focus first on the \emph{meaningful} queries, i.e., the queries without the randomness that is added to satisfy the security constraint. From the first database, the user asks for an individual symbol from every message, i.e., asks for $a_1,b_1,c_1$. From database 2, the user does not ask for new individual symbols but rather exploits the side information that is generated from database 1 to query for 2-sums from database 2, i.e., the user asks for $a_2+b_1$, $a_3+c_1$, $b_2+c_2$ from database 2. Then, the user exploits $b_2+c_2$ as side information to ask for $a_4+b_2+c_2$ from database 1. To get an integer number of downloads for the meaningful queries, which covers $(1-\mu_n)t_n$ from the downloaded symbols from the $n$th database, the scheme is repeated $\nu$ times. Since this scheme gets 4 symbols
from database 1 and 3 symbols from database 2, we choose the repetition factor of the scheme $\nu$ such that:
\begin{align}
(1-\mu_1)t_1=4\nu \:&\Rightarrow\: t_1=\frac{16\nu}{3}\\
(1-\mu_2)t_2=3\nu \:&\Rightarrow\: t_2=6\nu
\end{align} 
Then, the minimal $\nu$ is $\nu=3$. Database 1 generates the independent keys $K_1=\left(k_{1}^{(1)},\cdots, k_4^{(1)}\right)$, such that $K_1$ is picked uniformly from $\mathbb{F}_q^4$. Database 1 encodes these random keys using a $(16,4)$ MDS code, to get $u_{[1:16]}$, i.e., 
\begin{align}
u_{[1:16]}=\mds_{16 \times 4} K_1
\end{align}

Similarly, database 2 generates $K_2=\left(k_{1}^{(2)},\cdots, k_9^{(2)}\right)$ uniformly from $\mathbb{F}_q^9$. Database 2 encodes the keys using an $(18,9)$ MDS code, to get $v_{[1:16]}$, i.e.,
\begin{align}
v_{[1:18]}=\mds_{18 \times 9} K_2
\end{align}

Now, all the meaningful downloads are \emph{encrypted} by the coded keys. Furthermore, the user downloads $u_{[13:16]}$ separately from database 1, and $v_{[10:18]}$ from database 2. The query table is shown in Table.~\ref{M3N2_mu1_1/4_mu2_1/2}. 

\begin{table}[h]
	\centering
	\caption{The query table for $M=3$, $N=2$, $\mu_1=\frac{1}{4}$, $\mu_2=\frac{1}{2}$.}
	\label{M3N2_mu1_1/4_mu2_1/2}
	\begin{tabular}{|c|c|}
		\hline
		Database 1 & Database 2 \\
		\hline
		$a_1+u_1$  & $a_2+b_1+v_1$\\
		$b_1+u_2$  & $a_3+c_1+v_2$\\
		$c_1+u_3$  & $b_2+c_2+v_3$\\
		\hline
		$a_4+b_2+c_2+u_4$ &       \\
		\hline
		\hline
		$a_5+u_5$  & $a_6+b_3+v_4$\\
		$b_3+u_6$  & $a_7+c_3+v_5$\\
		$c_3+u_7$  & $b_4+c_4+v_6$\\
		\hline
		$a_8+b_4+c_4+u_8$ &       \\
		\hline
		\hline
		$a_9+u_9$  & $a_{10}+b_5+v_7$\\
		$b_5+u_{10}$  & $a_{11}+c_5+v_8$\\
		$c_5+u_{11}$  & $b_6+c_6+v_9$\\
		\hline
		$a_{12}+b_6+c_6+u_{12}$ &       \\
		\hline
		\hline
		$u_{13},u_{14},u_{15},u_{16}$ & $v_{10},u_{11},u_{12},u_{13},v_{14}$ \\
		                              & $v_{15},u_{16},u_{17},u_{18}$\\
		\hline
	\end{tabular}
\end{table}

For the decodability, since database 1 encodes its keys $K_1$ using a $(16,4)$ MDS code, by the MDS property, any $4$ symbols suffice to reconstruct $u_{[1:16]}$. The user downloads $u_{[13:16]}$ separately, hence $u_{[1:12]}$ can be reconstructed and canceled from the downloads to get the meaningful information only. Similarly, database 2 encodes the keys $K_2$ using an $(18,9)$ MDS code, hence $v_{[10:18]}$ suffice to reconstruct $v_{[1:9]}$ and can be canceled from the meaningful downloads. Furthermore, since the side information at any database is obtained from the undesired symbols downloaded from the second database, all undesired symbols can be canceled and the user is left only with $a_{[1:12]}$, which are the desired symbols.

For the security, since $\mu_1=\frac{1}{4}$ and $\mu_2=\frac{1}{2}$, the eavesdropper can obtain any $4$ symbols out of total 16 downloaded symbols from database 1, and any $9$ symbols out of total 18 downloaded symbols from database 2. Since $K_1$, $K_2$ are generated uniformly and independently from $\mathbb{F}_q^{4}$, $\mathbb{F}_q^{9}$, respectively, any $4$ symbols $(u_{i_1}, \cdots, u_{i_4})$ from $u_{[1:16]}$ are independent and uniformly distributed over $\mathbb{F}_q$, and similarly for any $9$ symbols $(v_{j_1}, \cdots, v_{j_9})$ from $v_{[1:18]}$. Consequently, the leakage at the eavesdropper is upper bounded by:
\begin{align}
I(W_{1:3};Z_{1:2}^{[1]})&=H(Z_{1:2})-H(Z_{1:2}|W_{1:3}) \\
                        &\leq \log_q 13-H\left(\begin{bmatrix}
                        u_{i_1}\\\vdots\\u_{i_4}\\v_{j_1} \\ \vdots\\ v_{j_9}
                        \end{bmatrix}\right)=0
\end{align} 

For the privacy, as all combinations of the sums are included in the queries and the indices of the message symbols are uniformly and independently permuted, the privacy constraint is satisfied. Hence, the user downloads $t_1=16$ symbols from database 1, and $t_2=18$ symbols from database 2. From these downloads, the user can decode $L=12$ symbols from $W_1$. Hence, $R=\frac{12}{34}=\frac{6}{17}$, which matches the upper bound. 

\subsubsection{Achieving the Upper Bound for Arbitrary $\bmu$}
Now, we show how to achieve the upper bound in \eqref{ub32} for general $\bmu$. As shown in the example of $\mu_1=\frac{1}{4}$, $\mu_2=\frac{1}{2}$, the user downloads $\mu_1 t_1$ as individual symbols from the coded keys from database 1, and $\mu_2 t_2$ as individual symbols from the coded keys from database 2. This leaves $(1-\mu_1)t_1$, $(1-\mu_2)t_2$, respectively for meaningful symbols. Furthermore, each scheme should be repeated $\nu$ times to ensure that $t_1, \,t_2 \in \mathbb{N}$. In the following, we focus on the meaningful symbols without the coded keys. We show only one repetition of the scheme.
\paragraph{For $R(\bmu)=\frac{1-\mu_1}{3}$:} To achieve this rate, the user applies the trivial retrieval scheme \cite{ChorPIR}, and downloads all messages from database 1, i.e., the user downloads $a_1,b_1,c_1$ from database 1. Hence, $t_2=0$ and
\begin{align}
(1-\mu_1)t_1=3\nu \: \Rightarrow \: t_1=\frac{3\nu}{1-\mu_1}
\end{align}
where $\nu$ is chosen such that $t_1 \in \mathbb{N}$. From every repetition, the user gets 1 symbol from $W_1$. Hence, $L=\nu$. The user asks for $\mu_1 t_1=\frac{3\mu_1\nu}{1-\mu_1}$ individual coded symbols from the keys, and the database encrypts the downloads with coded keys constructed from a  $(\frac{3\nu}{1-\mu_1},\frac{3\mu_1\nu}{1-\mu_1})$ MDS code. This ensures the security. The achievable rate in this case is
\begin{align}
R=\frac{L}{t_1+t_2}=\frac{\nu}{\frac{3\nu}{1-\mu_1}}=\frac{1-\mu_1}{3}
\end{align}

\paragraph{For $R(\bmu)=\frac{2(1-\mu_1)(1-\mu_2)}{3(1-\mu_2)+(1-\mu_1)}$:} To achieve this rate, the user downloads individual symbols from all messages from database 1, i.e., the user downloads $a_1,b_1,c_1$ from database 1. The user combines the two undesired symbols $b_1$, $c_1$ into a 2-sum $b_1+c_1$ and uses it as a side information in database 2. The query table for one repetition of the scheme for the meaningful symbols (without showing the keys) is shown in Table~\ref{M3N2corner2}.

\begin{table}[h]
	\centering
	\caption{The meaningful symbols for $M=3$, $N=2$ to achieve $\frac{2(1-\mu_1)(1-\mu_2)}{3(1-\mu_2)+(1-\mu_1)}$.}
	\label{M3N2corner2}
	\begin{tabular}{|c|c|}
		\hline
		Database 1 & Database 2 \\
		\hline
		$a_1,b_1,c_1$ & $a_2+b_1+c_1$\\
		\hline
	\end{tabular}
\end{table}

In this case, the scheme is repeated $\nu$ times such that $t_1,\,t_2 \in \mathbb{N}$,
\begin{align}
(1-\mu_1)t_1=3\nu \:&\Rightarrow\: t_1=\frac{3\nu}{1-\mu_1}\\
(1-\mu_2)t_2=1\nu \:&\Rightarrow\: t_2=\frac{\nu}{1-\mu_2}
\end{align}

Database 1 encodes $\mu_1t_1=\frac{3\nu\mu_1}{1-\mu_1}$ independent and uniformly distributed keys using a $(\frac{3\nu}{1-\mu_1},\frac{3\nu\mu_1}{1-\mu_1})$ MDS code to obtain the coded keys that are added to each download. Similarly, database 2 encodes  $\mu_2t_2=\frac{\nu\mu_2}{1-\mu_2}$ keys using a $(\frac{\nu}{1-\mu_2},\frac{\nu\mu_2}{1-\mu_2})$ MDS code to obtain the coded symbols. Using this scheme, the user decodes $L=2\nu$ from the desired messages. Consequently,
\begin{align}
R=\frac{L}{t_1+t_2}=\frac{2\nu}{\frac{3\nu}{1-\mu_1}+\frac{\nu}{1-\mu_2}}=\frac{2(1-\mu_1)(1-\mu_2)}{3(1-\mu_2)+(1-\mu_1)}
\end{align} 

\paragraph{For $R(\bmu)=\frac{4(1-\mu_1)(1-\mu_2)}{4(1-\mu_2)+3(1-\mu_1)}$:} An instance for this scheme is the $\mu_1=\frac{1}{4}$, $\mu_2=\frac{1}{2}$ example. To avoid repetition, we give only the general rate. As shown in the example, $t_1=\frac{4\nu}{1-\mu_1}$, and $t_2=\frac{3\nu}{1-\mu_2}$. From every repetition, the user can decode $4$ symbols, hence $L=4\nu$. Thus,
\begin{align}
R=\frac{L}{t_1+t_2}=\frac{4\nu}{\frac{4\nu}{1-\mu_1}+\frac{3\nu}{1-\mu_2}}=\frac{4(1-\mu_1)(1-\mu_2)}{4(1-\mu_2)+3(1-\mu_1)}
\end{align}

\begin{figure}[t]
	\centering
	\includegraphics[width=0.8\textwidth]{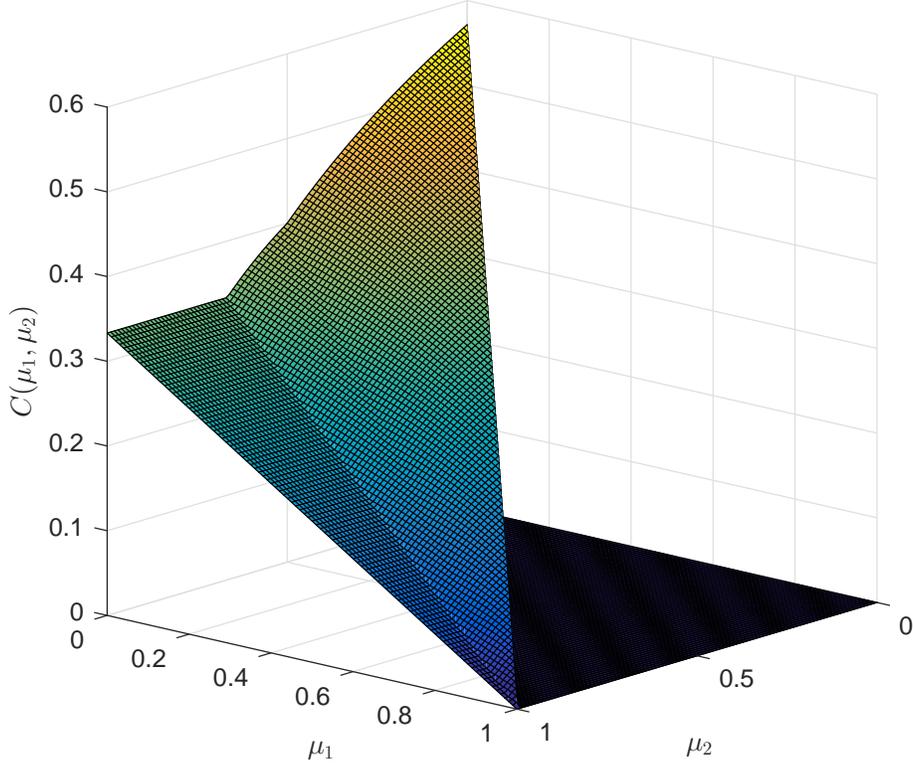}
	\caption{Capacity for $M=3$, $N=2$ as a function of $\mu_1$ and $\mu_2$.}
	\label{Fig: CapacityM3N2}
	\vspace*{-0.4cm}
\end{figure}

\begin{figure}[t]
	\centering
	\includegraphics[width=0.8\textwidth]{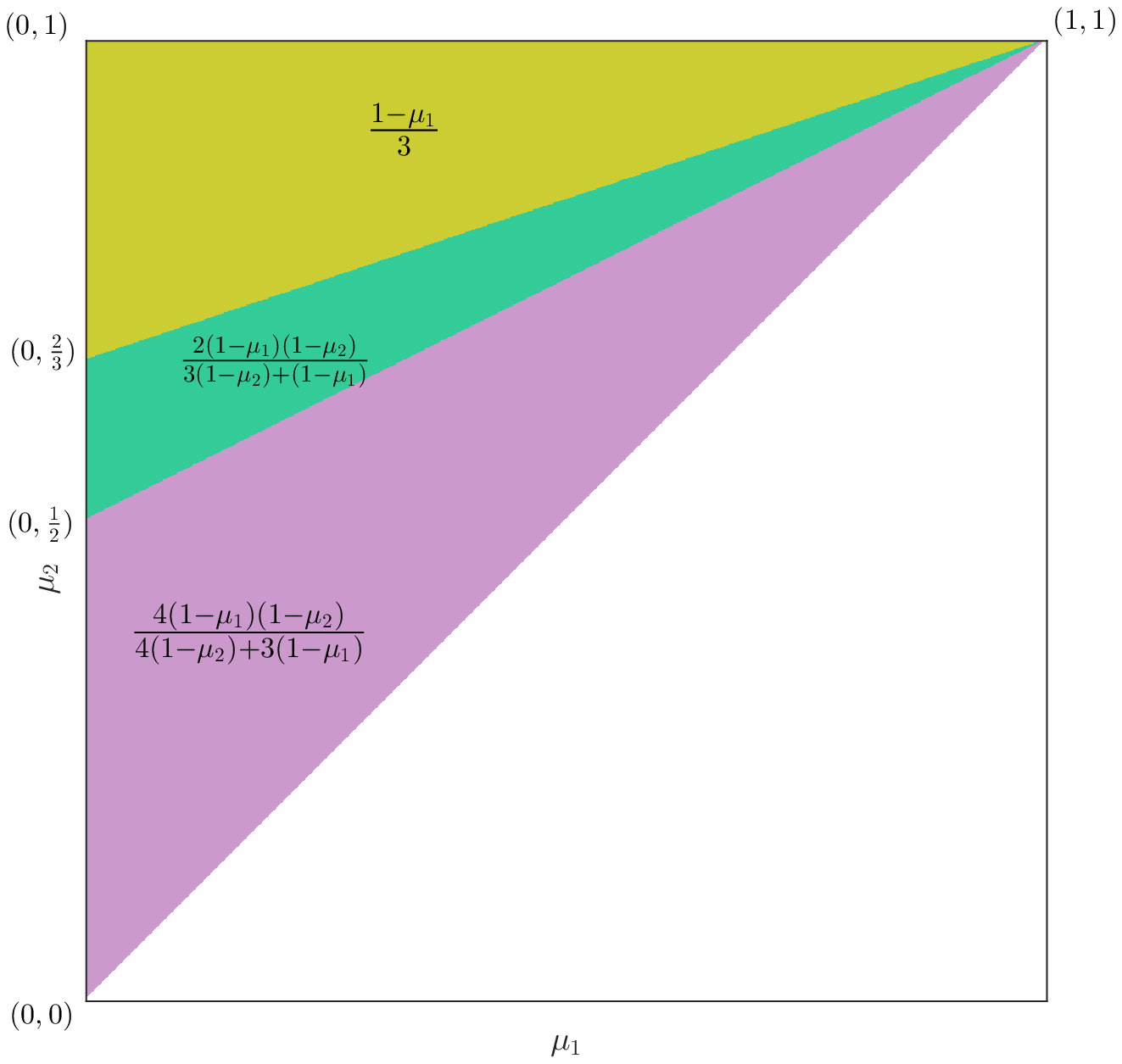}
	\caption{Partitions of $\bmu$ space according to the active capacity expression for $M=3$, $N=2$.}
	\label{Fig: RegionsM3N2}
	\vspace*{-0.4cm}
\end{figure}

This completes the description of the capacity-achieving scheme for PIR-WTC-II for $M=3$, $N=2$, and arbitrary $\bmu$. The capacity region $C(\bmu)$ is shown in Fig.~\ref{Fig: CapacityM3N2}. In Fig.~\ref{Fig: RegionsM3N2}, we illustrate the partitioning of the $\bmu$ space in terms of the active capacity expression; note by convention $\mu_2 \geq \mu_1$.
 
\subsection{General Achievable Scheme}
In this section, we present the general achievable scheme for PIR-WTC-II that achieves the retrieval rate in Theorem~\ref{Thm2}. The core of the achievable scheme is the achievable scheme of the corner points in the PIR problem under asymmetric traffic constraints in \cite{KarimAsymmetricPIR}. A new ingredient is needed to satisfy the security constraint, namely, encrypting the answer strings by random keys. The $n$th database uses a random key $K_n$ of length $\mu_n t_n$ that is sufficient to span the space of the eavesdropper's observations. The $n$th database encodes $K_n$ using a $(t_n,\mu_n t_n)$ MDS code and uses the resulting codeword to \emph{encrypt} each downloaded symbol from the meaningful downloads in addition to $\mu_n t_n$ individual symbols of coded key symbols only. For completeness, we include all related details of the scheme in \cite{KarimAsymmetricPIR} in addition to the new ingredients.

We use the same terminology as in \cite{KarimAsymmetricPIR}. Let $s_n \in \{0, 1, \cdots, M-1\}$ denote the number of side information symbols that are used simultaneously in the initial round of downloads at the $n$th database. For a given non-decreasing sequence $\{n_i\}_{i=0}^{M-1} \subset \{1, \cdots, N\}^M$, the databases are divided into groups, such that group 0 contains database 1 through database $n_0$, group 1 contains $n_1-n_0$ databases starting from database $n_0+1$, and so on.

Hence, let $s_n=i$ for all $n_{i-1}+1 \leq n \leq n_i$ with $n_{-1}=0$ by convention. Denote $\cs=\{i: s_n=i \:\text{for some}\: n \in \{1, \cdots, N\} \}$. We follow the round and stage definitions in \cite{MPIRjournal}. The $k$th round is the download queries that admit a sum of $k$ different messages ($k$-sum in \cite{JafarPIR}). A stage of the $k$th round is a query block of the $k$th round that exhausts all $\binom{M}{k}$ combinations of the $k$-sum. Denote $y_\ell[k]$ to be the number of stages in round $k$ downloaded from the $n$th database, such that $n_{\ell-1}+1 \leq n \leq n_\ell$. The details of the achievable scheme are as follows:

\begin{enumerate}
	\item \emph{Calculation of the number of repetitions:} The user and the databases agree on appropriate answer string lengths $t_n(\mathbf{n},\bmu)$, $n=1, \cdots, N$. To that end, the scheme associated with $\mathbf{n}=\{n_i\}_{i=0}^{M-1}$ is repeated $\nu$ times such that:
	\begin{align}\label{answer_length}
	t_n(\mathbf{n},\bmu)=\frac{\nu D_n(\mathbf{n})}{1-\mu_n}\: \in \mathbb{N},\quad \forall n \in \{1, \cdots, N\}
	\end{align}
	where $D_n(\mathbf{n})$ is the number of meaningful downloads corresponding to one repetition of the achievable scheme associated with the monotone non-decreasing sequence $\mathbf{n}=\{n_i\}_{i=0}^{M-1}$.
	\item \emph{Preparation of the keys:} The $n$th database generates a random key $K_n$. The random key $K_n$ is of length $\mu_n t_n$, such that elements of $K_n$ are independent and uniformly distributed over $\mathbb{F}_q$. The $n$th database encodes $K_n$ to an \emph{artificial noise} vector $u_{[1:t_n]}^{(n)}$ using a $(t_n,\mu_n t_n)$ MDS code, i.e., 
	\begin{align}
	u_{[1:t_n]}^{(n)}=\mds_{t_n \times \mu_nt_n} K_n
	\end{align}
	\item \emph{Initialization at the user side:} The user permutes each message independently and uniformly using a random interleaver, i.e., 
	\begin{align}
	x_m(i)=W_m(\pi_m(i)), \quad i \in \{1, \cdots, L\}
	\end{align}
	where $x_m(i)$ is the $i$th symbol of the permuted $W_m$, $\pi_m(\cdot)$ is a random interleaver for the $m$th message that is chosen independently, uniformly, and privately at the user's side. 
	\item \emph{Initial download:} From the $n$th database where $1 \leq n \leq n_0$, the user downloads $\prod_{s \in \cs} \binom{M-2}{s-1}$ symbols from the desired message. The user sets the round index $k=1$. I.e., the user starts downloading the desired symbols from $y_0[1]=\prod_{s \in \cs} \binom{M-2}{s-1}$ different stages.
	
	\item \emph{Message symmetry:} To satisfy the privacy constraint, for each stage initiated in the previous step, the user completes the stage by downloading the remaining $\binom{M-1}{k-1}$ $k$-sum combinations that do not include the desired symbols, in particular, if $k=1$, the user downloads $\prod_{s \in \cs} \binom{M-2}{s-1}$ individual symbols from each undesired message.
	
	\item \emph{Database symmetry:} We divide the databases into groups. Group $\ell \in \cs$ corresponds to databases $n_{\ell-1}+1$ to $n_{\ell}$. Database symmetry is applied within each group only. Consequently, the user repeats step~2 over each group of databases, in particular, if $k=1$, the user downloads $\prod_{s \in \cs} \binom{M-2}{s-1}$ individual symbols from each message from the first $n_0$ databases (group 1).
	
	\item \emph{Exploitation of side information:} The initial exploitation of side information is group-dependent as well. Specifically, the undesired symbols  downloaded within the $k$th round (the $k$-sums that do not include the desired message) are used as side information in the $(k+1)$th round. This exploitation of side information is performed by downloading $(k+1)$-sum consisting of 1 desired symbol and a $k$-sum of undesired symbols only that were generated in the $k$th round. However, the main difference from \cite{JafarPIR} is that, for the $n$th database, if $s_n>k$, then this database does not exploit the side information generated in the $k$th round. Consequently, the $n$th database belonging to the $\ell$th group exploits the side information generated in the $k$th round from all databases except itself if $s_n \leq k$. Moreover, for $s_n=k$, extra side information can be used in the $n$th database. This is due to the fact that the user can form $n_0\prod_{s \in \cs\setminus \{s_n\}} \binom{M-2}{s-1}$ extra stages of side information by constructing $k$-sums of the undesired symbols in round 1 from the databases in group 0. 
	
	\item \emph{Repeat} steps 5, 6, 7 after setting $k=k+1$ until $k=M$. 
	
	\item \emph{Repetition of the scheme:} Repeat steps $4,\cdots,8$ for a total of $\nu$ repetitions. 
	
	\item \emph{Shuffling the order of the queries:} By shuffling the order of the queries uniformly, all possible queries can be made equally likely regardless of the message index. This guarantees the privacy.
	\item \emph{Encryption of the downloads:} The database encrypts each meaningful download by adding one symbol from $u_{[1:(1-\mu_n)t_n]}^{(n)}$. Furthermore, the user downloads $u^{(n)}_{[(1-\mu_n)t_n+1:t_n]}$ coded key symbols individually. This guarantees the security.       
\end{enumerate}  

\subsection{Decodability, Privacy, Security, and Achievable Rate}
\paragraph{Decodability:} To see the decodability, we note that the user receives $\mu_n t_n$ individual artificial noise symbols $u^{(n)}_{[(1-\mu_n)t_n+1:t_n]}$ from the $n$th database. From the MDS property of the $(t_n,\mu_nt_n)$ MDS code, any $\mu_n t_n$ coded symbols suffice to reconstruct the entire $t_n$ coded symbols. Hence, the user can reconstruct and cancel $u^{(n)}_{[1:t_n]}$ by the knowledge of $u^{(n)}_{[(1-\mu_n)t_n+1:t_n]}$. Consequently, after canceling the artificial noise symbols, the user is left with only the meaningful symbols in the answer strings.

Now, by construction, in the $(k+1)$th round at the $n$th database, the user exploits the side information generated in the $k$th round in the remaining active databases by adding 1 symbol of the desired message with $k$-sum of undesired messages which was downloaded previously in the $k$th round. Moreover, for the $n$th database belonging to the $\ell$th group at the $(\ell+1)$th round, the user adds every $\ell$ symbols of the undesired symbols downloaded from group 0 to make one side information symbol. Since the user downloads $\prod_{\ell \in \cs} \binom{M-2}{\ell-1}$ from every database in the first $n_0$ databases (group 0), the user can exploit such side information to initiate $n_0 \prod_{\ell \in \cs \setminus \{\ell\}} \binom{M-2}{\ell-1}$ stages in the $(\ell+1)$th round from every database in group $\ell$. Since all side information symbols used in the $(k+1)$th round is decodable in the $k$th round or from round 1, the user cancels out these side information and is left with symbols from the desired message. 

\paragraph{Privacy:} The privacy of the scheme follows from the privacy of the inherent PIR scheme under asymmetric traffic constraints. Specifically, for every stage of the $k$th round initiated in the exploitation of the side information step, all $\binom{M}{k}$ combinations of the $k$-sum are included at each round. Thus, the structure of the queries is the same for any desired message. The privacy constraint in \eqref{privacy_constraint} is satisfied by the random and independent permutation of each message and the random shuffling of the order of the queries. This ensures that all queries are equally likely independent of the desired message index.

\paragraph{Security:} From the $n$th database key $K_n$ is of length $\mu_n t_n$. The elements of $K_n$ are independent and uniformly distributed in $\mathbb{F}_q$. The $n$th database encodes $K_n$ into the artificial noise vector $u_{[1:t_n]}^{(n)}$ using a $(t_n, \mu_n t_n)$ MDS code. Since any $\mu_n t_n$ columns of the generator matrix of the MDS code are full rank, the mapping from $K_n$ to any $\mu_n t_n$ symbols from the artificial noise vector $U_n=[u_{i_1}^{(n)}, \cdots, u_{i_{\mu_n t_n}}^{(n)}]$ is a bijection, and consequently, $U_n \sim K_n$, where $\sim$ denotes statistical equivalence. Moreover, since there is no shared randomness between databases, the elements of $(K_1, \cdots, K_N)$, and consequently the elements of $(U_1, \cdots, U_N)$ are independent and uniformly distributed in $\mathbb{F}_q$.

Now, the eavesdropper chooses to observe $\mu_n t_n$ symbols from the $n$th answer string $A_n^{[i]}$. Denote the eavesdropper observations by $Z_n^{[i]} \in \mathbb{F}_q^{\mu_n t_n}$. Since all downloaded symbols are encrypted using $u_{[1:t_n]}^{(n)}$ (counting the downloads that contain solely the artificial noise). Denote the artificial noise symbols within $Z_n^{[i]}$ by $U_n$. Hence, the leakage at the eavesdropper can be upper bounded by:
\begin{align}
I(W_{1:M};Z_{1:N}^{[i]})&=H(Z_{1:N}^{[i]})-H(Z_{1:N}^{[i]}|W_{1:M})\\
                        &\leq \sum_{n=1}^N \mu_n t_n-H\left(\begin{bmatrix}
                        U_1\\U_2\\ \vdots\\ U_N
                        \end{bmatrix}\right)\\
                        &=\sum_{n=1}^N \mu_n t_n-\sum_{n=1}^N \mu_n t_n=0 \label{security}
\end{align}
where \eqref{security} follows from the fact that any $\mu_n t_n$ artificial noise symbols are independent. Note that the units of calculation is $q$-ary symbols.

\paragraph{Achievable Rate:} For the calculation of the achievable rate, we focus first on one repetition of the scheme. Without adding the artificial noise symbols, the structure of one repetition of our scheme is exactly as \cite{KarimAsymmetricPIR}. The recursive structure of the achievable scheme can be described using the following system of difference equations that relate the number of stages in the databases belonging to a specific group as shown in \cite[Theorem~2]{KarimAsymmetricPIR}:
	\begin{align}\label{difference_eqn}
	y_0[k]&=(n_0\!-\!1)y_0[k\!-\!1]+\sum_{j \in \cs \setminus \{0\}} (n_j\!-\!n_{j-1}) y_j[k\!-\!1] \notag\\
	y_1[k]&=(n_1\!-\!n_0\!-\!1)y_1[k\!-\!1]+\sum_{j \in \cs \setminus \{1\}} (n_j\!-\!n_{j-1}) y_j[k\!-\!1] \notag\\
	y_\ell[k]&=n_0 \xi_\ell \delta[k\!-\!\ell\!-\!1]+(n_\ell\!-\!n_{\ell-1}\!-\!1) y_\ell[k-1]+\sum_{j \in \cs \setminus \{\ell\}} (n_j\!-\!n_{j-1})y_j[k\!-\!1], \quad  \ell \geq 2
	\end{align}
where $y_\ell[k]$ is the number of stages in the $k$th round in a database belonging to the $\ell$th group, i.e., for the $n$th database, such that $n_{\ell-1}+1 \leq n \leq n_\ell$.

Hence, to calculate $D_n(\mathbf{n})$ such that $n_{\ell-1} \leq n \leq n_{\ell}$, which is the number of meaningful downloads from the $n$th database belonging to the $\ell$th group, corresponding to one repetition of the achievable scheme associated with the sequence $\mathbf{n}=\{n_i\}_{i=0}^{M-1}$, we note that for any stage in the $k$th round, the user downloads $\binom{M-1}{k-1}$ desired symbols from a total of $\binom{M}{k}$ downloads. Therefore,
\begin{align}
D_n(\mathbf{n})=\sum_{k=1}^{M} \binom{M}{k} y_\ell[k], \quad n_{\ell-1} \leq n \leq n_{\ell} 
\end{align}

Consequently, the total download $\sum_{n=1}^N t_n(\mathbf{n})$ from all databases from all repetitions is calculated by observing \eqref{answer_length},
\begin{align}
\sum_{n=1}^N t_n(\mathbf{n},\bmu)&= \sum_{n=1}^N\frac{\nu D_n(\mathbf{n})}{1-\mu_n}\\
                            &=\nu\left[\sum_{n=1}^{n_0}\frac{\sum_{k=1}^{M} \binom{M}{k} y_0[k] }{1-\mu_n}+\sum_{n=n_0+1}^{n_1}\frac{\sum_{k=1}^{M} \binom{M}{k} y_1[k]}{1-\mu_n}+\cdots\right]\\
                            &= \nu\sum_{\ell \in \cs} \sum_{n=n_{\ell-1}+1}^{n_\ell}\frac{\sum_{k=1}^{M} \binom{M}{k} y_\ell[k] }{1-\mu_n}
\end{align}
Furthermore, the total desired symbols from all databases from all repetitions is given by,
\begin{align}
L(\mathbf{n})=\nu\sum_{\ell \in \cs} \sum_{k=1}^{M}\binom{M-1}{k-1} y_\ell[k](n_\ell-n_{\ell-1})
\end{align}
Thus, the following rate is achievable corresponding to the sequence $\mathbf{n}$,
\begin{align}
R(\mathbf{n},\bmu)=\frac{\sum_{\ell \in \cs} \sum_{k=1}^{M}\binom{M-1}{k-1} y_\ell[k](n_\ell-n_{\ell-1})}{\sum_{\ell \in \cs} \sum_{n=n_{\ell-1}+1}^{n_\ell}\frac{\sum_{k=1}^{M} \binom{M}{k} y_\ell[k] }{1-\mu_n}}
\end{align}

Since this scheme is achievable for every monotone non-decreasing sequence $\mathbf{n}=\{n_i\}_{i=0}^{M-1}$, the following rate is achievable,
\begin{align}
R(\bmu)= \max_{n_0 \leq \cdots \leq n_{M-1} \in \{1, \cdots, N\}} \frac{\sum_{\ell \in \cs} \sum_{k=1}^{M}\binom{M-1}{k-1} y_\ell[k](n_\ell-n_{\ell-1})}{\sum_{\ell \in \cs} \sum_{n=n_{\ell-1}+1}^{n_\ell}\frac{\sum_{k=1}^{M} \binom{M}{k} y_\ell[k] }{1-\mu_n}}
\end{align} 

\subsection{Optimality for $M=2$ and $M=3$ Messages}\label{proofM3N2}
In this section, we prove the optimality of our scheme for $M=2$ and $M=3$. The proof relies on relating the upper bound for the PIR-WTC-II problem with the upper bound for the PIR problem under asymmetric traffic constraints. From the settled optimality of the achievable scheme of the meaningful symbols for $M=2$, $M=3$ for the PIR problem under asymmetric traffic constraints, we conclude the optimality of our scheme for PIR-WTC-II.\footnote{Alternatively, for a specified $N$, $\bmu$, we can prove the optimality by showing that the KKT conditions of the upper bound optimization problem are satisfied by our achievable scheme.}

We return to the upper bound in Theorem~\ref{Thm1},
\begin{align}
\bar{C}(\bmu)&=\max_{\boldsymbol{\tau} \in \mathbb{T}} \min_{n_i \in \{1, \cdots, N\}} \frac{\sum_{n=1}^N (1-\mu_n)\tau_n+\frac{\sum_{n=n_1+1}^N (1-\mu_n)\tau_n}{n_1}+\cdots+\frac{\sum_{n=n_{M-1}+1}^N (1-\mu_n)\tau_n}{\prod_{i=1}^{M-1} n_i}}{1+\frac{1}{n_1}+\cdots+\frac{1}{\prod_{i=1}^{M-1}n_i}}\\
             &=\max_{\boldsymbol{\tau} \in \mathbb{T}} \sum_{n=1}^N (1-\mu_n)\tau_n \cdot \!\!\min_{n_i \in \{1, \cdots, N\}} \frac{1+\frac{\sum_{n=n_1+1}^N (1-\mu_n)\tau_n}{n_1\cdot \sum_{n=1}^N (1-\mu_n)\tau_n}+\cdots+\frac{\sum_{n=n_{M-1}+1}^N (1-\mu_n)\tau_n}{\prod_{i=1}^{M-1} n_i\cdot \sum_{n=1}^N (1-\mu_n)\tau_n}}{1+\frac{1}{n_1}+\cdots+\frac{1}{\prod_{i=1}^{M-1}n_i}}\\
             &=\max_{\boldsymbol{\tau} \in \mathbb{T}} \sum_{n=1}^N (1-\mu_n)\tau_n \cdot \!\!\!\!\min_{n_i \in \{1, \cdots, N\}} \frac{1+\frac{1}{n_1}\sum_{n=n_1+1}^N\!\! \tilde{\tau}_n+\cdots+\frac{1}{\prod_{i=1}^{M-1}n_i}\sum_{n=n_{M-1}+1}^N\!\! \tilde{\tau}_n}{1+\frac{1}{n_1}+\cdots+\frac{1}{\prod_{i=1}^{M-1}n_i}}\label{ub_LP}\\
             &=\max_{\boldsymbol{\tau} \in \mathbb{T}} \sum_{n=1}^N (1-\mu_n)\tau_n \cdot \tilde{C}(\tilde{\bt}) \label{inner-outer}
\end{align}
where $\tilde{\tau}_n$ is obtained by the change of variable $\tilde{\tau}_n=\frac{(1-\mu_n)\tau_n}{\sum_{i=1}^N (1-\mu_i)\tau_i}$ and the inner problem $\tilde{C}(\tilde{\bt})$ is defined as:
\begin{align}
\tilde{C}(\tilde{\bt})=\min_{n_i \in \{1, \cdots, N\}} \frac{1+\frac{1}{n_1}\sum_{n=n_1+1}^N\!\! \tilde{\tau}_n+\cdots+\frac{1}{\prod_{i=1}^{M-1}n_i}\sum_{n=n_{M-1}+1}^N\!\! \tilde{\tau}_n}{1+\frac{1}{n_1}+\cdots+\frac{1}{\prod_{i=1}^{M-1}n_i}}
\end{align}
The inner problem is precisely the upper bound for the PIR problem under asymmetric traffic constraints $\tilde{\bt}$ in \cite[Theorem~1]{KarimAsymmetricPIR}.

In the following lemma, we show that the solution of $\bar{C}(\bmu)$ exists at one of the corner points of $\tilde{C}(\tilde{\bt})$.

\begin{lemma}
	The solution of $\bar{C}(\bmu)$ exists at one of the corner points of $\tilde{C}(\tilde{\bt})$ after the change of variables $\tau_n=\frac{\sum_{i=1}^N (1-\mu_i)\tau_i}{(1-\mu_n)}$.
\end{lemma}

\begin{Proof}
To show this, we note that the upper bound in Theorem~\ref{Thm1} can be written as the following linear program as discussed in Remark~\ref{Remark3}:
    \begin{align}\label{LP}
    \max_{\bt,R}  &\quad R \notag\\
    \st &\quad R \leq \frac{\sum_{n=1}^N (1-\mu_n)\tau_n+\frac{\sum_{n=n_1+1}^N (1-\mu_n)\tau_n}{n_1}+\cdots+\frac{\sum_{n=n_{M-1}+1}^N (1-\mu_n)\tau_n}{\prod_{i=1}^{M-1} n_i}}{1+\frac{1}{n_1}+\cdots+\frac{1}{\prod_{i=1}^{M-1}n_i}}, \quad \forall \mathbf{n}\notag\\
    &\quad \sum_{n=1}^{N}\tau_n=1,\quad \tau_n \geq 0, \: n=1, \cdots, N 
    \end{align}

Equivalently, from \eqref{ub_LP}, we can write the optimization problem corresponding to the upper bound as:

    \begin{align}\label{LP2}
    \max_{\bt \in \mathbb{T},\tilde{R},\tilde{\bt}}  &\quad \sum_{n=1}^N (1-\mu_n)\tau_n \cdot \tilde{R} \notag\\
    \st &\quad \tilde{R} \leq \frac{1+\frac{1}{n_1}\sum_{n=n_1+1}^N\!\! \tilde{\tau}_n+\cdots+\frac{1}{\prod_{i=1}^{M-1}n_i}\sum_{n=n_{M-1}+1}^N\!\! \tilde{\tau}_n}{1+\frac{1}{n_1}+\cdots+\frac{1}{\prod_{i=1}^{M-1}n_i}}, \quad \forall \mathbf{n}\notag\\
    &\quad \sum_{n=1}^{N} \tilde{\tau}_n=1,\quad \tilde{\tau}_n \geq 0, \: n=1, \cdots, N \notag\\
    &\quad \tilde{\tau}_n=\frac{(1-\mu_n)\tau_n}{\sum_{i} (1-\mu_i)\tau_i}, \quad n=1, \cdots, N
    \end{align}
We note that the constraints of this equivalent problem is the same as constraints of the upper bounds of the PIR problem under the asymmetric traffic constraints $\tilde{\bt}$. 

Since there are a finite number of constraints ($N^{M-1}+2$ constraints), the feasible region is a polyhedron, thus, the solution for $\bar{C}(\bmu)$ resides at a corner point of this polyhedron. 

For any corner point of this optimization problem, $(N+1)$ constraints are active (i.e., met with equality) and linearly independent. 

Since these constraints take the form of
\begin{align}
R = \frac{\sum_{n=1}^N (1-\mu_n)\tau_n+\frac{\sum_{n=n_1+1}^N (1-\mu_n)\tau_n}{n_1}+\cdots+\frac{\sum_{n=n_{M-1}+1}^N (1-\mu_n)\tau_n}{\prod_{i=1}^{M-1} n_i}}{1+\frac{1}{n_1}+\cdots+\frac{1}{\prod_{i=1}^{M-1}n_i}}
\end{align} 
by dividing both sides by $\sum_{i=1}^N (1-\mu_i)\tau_i > 0$, the constraint become 
\begin{align}
\tilde{R}=\frac{R}{\sum_{i=1}^N (1-\mu_i)\tau_i}=\frac{1+\frac{1}{n_1}\sum_{n=n_1+1}^N\!\! \tilde{\tau}_n+\cdots+\frac{1}{\prod_{i=1}^{M-1}n_i}\sum_{n=n_{M-1}+1}^N\!\! \tilde{\tau}_n}{1+\frac{1}{n_1}+\cdots+\frac{1}{\prod_{i=1}^{M-1}n_i}}
\end{align}
Hence, the condition of intersection of the active constraints of the $\bar{C}(\bmu)$ is the same as the condition of the intersection of the bounds of $\tilde{C}(\tilde{\bt})$ after the change of variables. Thus, it suffices to consider the corner points of the inner problem and map the solution using the change of variables $\tau_n=\frac{\sum_{i=1}^N (1-\mu_i)\tau_i}{(1-\mu_n)}$.
\end{Proof}

Consequently, for a corner point of the inner problem $(\tilde{\bt}^*,\tilde{C}(\tilde{\bt}^*))$, we have the reverse change of variables
\begin{align}
\tau_n^*=\tilde{\tau}_n^* \cdot\frac{\sum_{i=1}^{N} (1-\mu_i) \tau_i^*}{1-\mu_n}
\end{align}

Now, since $\sum_{n=1}^N \tau_n^*=1$, $\sum_{n=1}^N \tilde{\tau}_n^* \cdot\frac{\sum_{i=1}^{N} (1-\mu_i) \tau_i^*}{1-\mu_n}=1$, which leads to 
\begin{align}
	\sum_{i=1}^{N} (1-\mu_i) \tau_i=\frac{1}{\sum_{n=1}^{N} \frac{\tilde{\tau}_n}{1-\mu_n}}
\end{align}
Denote $\bar{C}(\tilde{\bt}^*,\bmu)$ to be the upper bound of the PIR-WTC-II problem corresponding to the corner point $(\tilde{\bt}^*,\tilde{C}(\tilde{\bt}^*))$ of the inner problem, hence from \eqref{inner-outer}, we have
\begin{align}
\bar{C}(\tilde{\bt}^*,\bmu)&=\sum_{i=1}^{N} (1-\mu_i) \tau_i\cdot \tilde{C}(\tilde{\bt}^*)\\
                           &=\frac{\tilde{C}(\tilde{\bt}^*)}{\sum_{n=1}^{N} \frac{\tilde{\tau}_n}{1-\mu_n}}
\end{align} 
Thus, the upper bound can be written in terms of the corner points of the inner problem $\{\tilde{\bt}^{(i)}\}_{i=1}^{\theta}$, where $\theta$ is the total number of corner points as 
\begin{align}\label{explicit_ub}
\bar{C}(\bmu)=\max_{i \in \{1, \cdots, \theta\}} \frac{\tilde{C}(\tilde{\bt}^{(i)})}{\sum_{n=1}^{N} \frac{\tilde{\bt}^{(i)}}{1-\mu_n}}
\end{align}

\subsubsection{$M=2$ Messages} From \cite{KarimAsymmetricPIR}, we know that for $M=2$, all the corner points of the inner problem are in fact optimal. For an increasing sequence $(n_0,n_1)$, the corner points are characterized by:
\begin{align}
\tilde{\tau}_{n}= 
\left\{
\begin{array}{ll}
\frac{n_0+1}{n_0(n_1+1)}, \quad &1 \leq n \leq n_0\\
\frac{1}{n_1+1}, \quad &n_0+1 \leq n \leq n_1\\
0, \quad &n_1+1 \leq n \leq N
\end{array}
\right. 
\quad \Rightarrow \quad \tilde{C}(\tilde{\bt})=
\frac{n_1}{n_1+1}
\end{align}
Hence, the upper bound for $M=2$ can be explicitly written as:
\begin{align}
\bar{C}(\bmu)&=\max_{n_0,n_1 \in \{1, \cdots, N\}}\frac{\frac{n_1}{n_1+1}}{\sum_{n=1}^{n_0} \frac{n_0+1}{n_0(n_1+1)(1-\mu_n)}+\sum_{n=n_0+1}^{n_1} \frac{1}{(n_1+1)(1-\mu_n)}}\\
             &=\max_{n_0,n_1 \in \{1, \cdots, N\}} \frac{n_0 n_1}{\sum_{n=1}^{n_0} \frac{n_0+1}{1-\mu_n}+\sum_{n=n_0+1}^{n_1} \frac{n_0}{1-\mu_n}}
\end{align}

From the achievability side, for a sequence $(n_0,n_1)$, the system of difference equations in Theorem~\ref{Thm2} reduces to
\begin{align}
y_0[k]&=(n_0-1)y_0[k-1] \\
y_1[k]&=n_0 y_0[k-1]
\end{align}
for $k=1,2$, where $y_0[1]=1$, and $y_1[1]=0$. Hence, $y_0[2]=n_0-1$, and $y_1[2]=n_0$. Consequently, the achievable rate in Theorem~\ref{Thm2} is explicitly evaluated for $M=2$ as:
\begin{align}
R(\bmu)&= \max_{n_0, n_1 \in \{1, \cdots, N\}} \frac{\sum_{\ell \in \cs} \sum_{k=1}^{M}\binom{M-1}{k-1} y_\ell[k](n_\ell-n_{\ell-1})}{\sum_{\ell \in \cs} \sum_{n=n_{\ell-1}+1}^{n_\ell}\frac{\sum_{k=1}^{M} \binom{M}{k} y_\ell[k] }{1-\mu_n}}\\
&=\max_{n_0,n_1 \in \{1, \cdots, N\}} \frac{n_0 n_1}{\sum_{n=1}^{n_0} \frac{n_0+1}{1-\mu_n}+\sum_{n=n_0+1}^{n_1} \frac{n_0}{1-\mu_n}}
\end{align} 
which matches the upper bound and concludes the optimality for $M=2$.
 
\subsubsection{$M=3$ Messages} Similarly, from \cite{KarimAsymmetricPIR}, the corner points of the inner problem occur for an increasing sequence $(n_0,n_1,n_2)$. The corner points are characterized by:
\begin{align}
\tilde{\tau}_{n}= 
\left\{
\begin{array}{ll}
\frac{n_0n_1+n_0+1}{n_0(n_2n_1+n_1+1)}, \quad &1 \leq n \leq n_0\\
\frac{n_1+1}{n_2n_1+n_1+1}, \quad &n_0+1 \leq n \leq n_1\\
\frac{n_1}{n_2n_1+n_1+1}\quad  &n_1+1 \leq n \leq n_2\\
0, \quad &n_2+1 \leq n \leq N
\end{array}
\right. 
\quad \Rightarrow \quad \tilde{C}(\tilde{\bt})=
\frac{n_1 n_2}{n_1n_2+n_1+1}
\end{align}
Hence, the upper bound in \eqref{explicit_ub} is explicitly written as:
\begin{align}
\bar{C}(\bmu)&= \max_{n_0,n_1,n_2 \in \{1, \cdots, N\}} \frac{n_0n_1n_2}{\sum_{n=1}^{n_0} \frac{n_0n_1+n_0+1}{1-\mu_n}+\sum_{n=n_0+1}^{n_1} \frac{n_0n_1+n_0}{1-\mu_n}+\sum_{n=n_1+1}^{n_2} \frac{n_0n_1}{1-\mu_n}}
\end{align}

From the achievability side, we have the following system of difference equations for $k=1,2,3$:
\begin{align}
y_0[k]&=(n_0-1)y_0[k-1]+(n_1-n_0)y_1[k-1]+(n_2-n_1)y_2[k-1] \\
y_1[k]&=n_0y_0[k-1]+(n_1-n_0-1)y_1[k-1]+(n_2-n_1)y_2[k-1] \\
y_2[k]&=n_0\delta[k-3]+n_0y_0[k-1]+(n_1-n_0)y_1[k-1]+(n_2-n_1-1)y_2[k-1]
\end{align}
with the initial conditions $y_0[1]=1$, $y_1[1]=0$, and $y_2[1]=y_2[2]=0$. Evaluating $y_\ell[k]$, for $\ell=0,1,2$, and $k=1,2,3$ recursively leads to $y_0[2]=n_0-1$, $y_1[2]=n_0$, $y_0[3]=n_1n_0-2n_0+1$, $y_1[3]=n_1n_0-2n_0$, and $y_2[3]=n_1n_0$. Consequently, the achievable rate from Theorem~\ref{Thm2} is explicitly expressed as:
\begin{align}
R(\bmu)&= \max_{n_0, n_1 \in \{1, \cdots, N\}} \frac{\sum_{\ell \in \cs} \sum_{k=1}^{M}\binom{M-1}{k-1} y_\ell[k](n_\ell-n_{\ell-1})}{\sum_{\ell \in \cs} \sum_{n=n_{\ell-1}+1}^{n_\ell}\frac{\sum_{k=1}^{M} \binom{M}{k} y_\ell[k] }{1-\mu_n}}\\
&=\max_{n_0,n_1,n_2 \in \{1, \cdots, N\}} \frac{n_0n_1n_2}{\sum_{n=1}^{n_0} \frac{n_0n_1+n_0+1}{1-\mu_n}+\sum_{n=n_0+1}^{n_1} \frac{n_0n_1+n_0}{1-\mu_n}+\sum_{n=n_1+1}^{n_2} \frac{n_0n_1}{1-\mu_n}}
\end{align}
which matches the upper bound and concludes the optimality for $M=3$.

\begin{remark}
	We note that the meaningful portion of the answer strings follows the combinatorial water-filling shown in \cite{KarimAsymmetricPIR} for $M=2$ and $M=3$. This means that the less threatened (more secure) databases are returning more meaningful symbols than the less secure ones, hence, $\tilde{\tau}_n \geq \tilde{\tau}_k$, if $n < k$. However, the length of the entire answer string including the artificial noise symbols may not follow the same structure, e.g., in the example in Section~\ref{example}, we see that $t_1=16$ and $t_2=18$, i.e., $\tau_2>\tau_1$, while $\tilde{\tau}_2 < \tilde{\tau}_1$.   
\end{remark}
\subsection{Achievable Rate for $N=2$ and Arbitrary $M$}\label{N2}
Following the analysis of this case in \cite{KarimAsymmetricPIR}, let $s_2 \in \{0, \cdots, M-1\}$ be the number of side information symbols that are used simultaneously in the initial round download in the second database.

Hence, the user starts with downloading $\binom{M-2}{s_2-1}$ stages of individual symbols (i.e., the user downloads $M\binom{M-2}{s_2-1}$ symbols from round 1 from all messages) from the first database to create 1 stage of side information in the $(s_2+1)$th round. After the initial exploitation of side information, the two databases exchange side information.  More specifically, from database 1 in the $(s_2+2k)$th round, where $k=1, \cdots, \left\lfloor \frac{M-s_2}{2}\right\rfloor$, the user exploits the side information generated in database 2 in the $(s_2+2k-1)$th round to download $\binom{M-1}{s_2+2k-1}$ desired symbols from total download in the $(s_2+2k)$th round of $\binom{M}{s_2+2k}$. Similarly from database 2, in the $(s_2+2k+1)$th round, where $k=0, \cdots, \left\lfloor \frac{M-s_2-1}{2}\right\rfloor$, the user exploits the side information generated in database 1 in the $(s_2+2k)$th round, and downloads $\binom{M-1}{s_2+2k}$ desired symbols from total of $\binom{M}{s_2+2k+1}$ downloads in the $(s_2+2k+1)$th round. Thus, using the calculation in \cite{KarimAsymmetricPIR}, we have
\begin{align}
D_1(s_2)&=M\binom{M-2}{s_2-1}+\sum_{k=1}^{\left\lfloor \frac{M-s_2}{2}\right\rfloor} \binom{M}{s_2+2k} \\
D_2(s_2)&=\sum_{k=0}^{\left\lfloor \frac{M-s_2-1}{2}\right\rfloor} \binom{M}{s_2+2k+1}
\end{align}
where $D_n(s_2)$ corresponds to the length of the meaningful downloads within the $n$th database from one repetition of the scheme, therefore, the total download of the scheme is given by:
\begin{align}
t_1(s_2)+t_2(s_1)=&\frac{D_1(s_2)}{1-\mu_1}+\frac{D_2(s_2)}{1-\mu_2}\\
                 =&\frac{1}{1-\mu_1} \left[M\binom{M-2}{s_2-1}
                 +\sum_{k=1}^{\left\lfloor \frac{M-s_2}{2}\right\rfloor} \binom{M}{s_2+2k} \right]\notag\\
                 &+\frac{1}{1-\mu_2}\left[\sum_{k=0}^{\left\lfloor \frac{M-s_2-1}{2}\right\rfloor} \binom{M}{s_2+2k+1}\right]
\end{align}
The message length does not change due to the security constraint, hence, directly from \cite{KarimAsymmetricPIR}, we have
\begin{align}
L(s_2)=\binom{M-2}{s_2-1}+\sum_{k=0}^{M-s_2-1}\binom{M-1}{s_2+k}
\end{align}
Consequently, the achievable rate is explicitly given as:
\begin{align}
R(\bmu)=\max_{s_2 \in \{0, \cdots, M-1\}}\frac{\binom{M-2}{s_2-1}+\sum_{k=0}^{M-s_2-1}\binom{M-1}{s_2+k}}{\frac{1}{1-\mu_1} \left[M\binom{M-2}{s_2-1}
	\!\!+\!\!\sum_{k=1}^{\left\lfloor \frac{M-s_2}{2}\right\rfloor} \binom{M}{s_2+2k} \right]\!+\!\frac{1}{1-\mu_2}\!\left[\sum_{k=0}^{\left\lfloor \frac{M-s_2-1}{2}\right\rfloor} \binom{M}{s_2+2k+1}\right]}
\end{align}
including the corner point corresponding to the trivial rate, i.e., when the user deactivates the retrieval process from the second database, leading to \eqref{achievableN2}.

\subsection{Further Examples}
In this section, we present further examples to clarify the achievable scheme for additional tractable values of $M$, $N$. 
\subsubsection{$M=4$ Messages, $N=2$ Databases}
In this example, we show the achievable scheme for $M=4$, $N=2$, and arbitrary $\bmu$. This example helps us to show that our achievable scheme does not achieve the capacity for all $\bmu$. For $M=4$, we have $M+1=5$ possible achievable schemes, corresponding to $s_2 =\{0,1, \cdots, 3\}$ and one other achievable scheme corresponding to the trivial scheme of downloading the contents of database 1. Let $a_i, b_i, c_i, d_i$ denote the randomly permuted symbols from $W_1, W_2, W_3, W_4$, respectively. In all achievable schemes, the $n$th database generates a key $K_n$ with length $\mu_n t_n$ and encodes it to generate an artificial noise vector $u_{[1:t_n]}^{(n)}$ using a $(t_n,\mu_n t_n)$ MDS code. The $n$th database provides $\mu_n t_n$ individual symbols of artificial noise. In all cases, the scheme is repeated $\nu$ times such that:
	\begin{align}\label{answer_lengthM4}
	t_n(\mathbf{n},\bmu)=\frac{\nu D_n(\mathbf{n})}{1-\mu_n}\: \in \mathbb{N},\quad \forall n \in \{1, 2\}
	\end{align}

Now, we focus on one repetition of the achievable scheme. We further concentrate on the meaningful queries, i.e., before adding the artificial noise vector. 

\paragraph{The trivial scheme corresponding to $\mathbf{n}=(1,1,1,1)$:}  In one repetition of the scheme, the user downloads $a_1, b_1, c_1, d_1$ from database 1. Hence, $D_1(\mathbf{n})=4$. Consequently, $t_1(\mathbf{n},\bmu)=\frac{4\nu}{1-\mu_1}$. As the user decodes $1$ symbol from $W_1$ in each repetition, $L_1(\mathbf{n})=\nu$. Hence, $R(\mathbf{n}, \bmu)=\frac{1-\mu_1}{4}$ is achievable.

\paragraph{The scheme corresponding to $\mathbf{n}=(1,1,1,2)$:} In this case, $s_2=3$, i.e., the user exploits $3$ side-information symbols simultaneously in database 2, i.e., focusing on one repetition of the scheme, from database 1, the user downloads $a_1,b_1,c_1,d_1$. The user combines $b_1+c_1+d_1$ and uses this side information to get $a_2$ from database 2, i.e., the user downloads $a_2+b_1+c_1+d_1$. Hence, $D_1(\mathbf{n})=4$, $D_2(\mathbf{n})=1$. Consequently, $t_1(\mathbf{n},\bmu)=\frac{4\nu}{1-\mu_1}$, and $t_2(\mathbf{n},\bmu)=\frac{\nu}{1-\mu_2}$. As the user decodes $2$ symbols from $W_1$ in each repetition, $L_1(\mathbf{n})=2\nu$. Hence, $R(\mathbf{n}, \bmu)=\frac{2}{\frac{4}{1-\mu_1}+\frac{1}{1-\mu_2}}$ is achievable. The query table of the meaningful queries (without the artificial noise) for one repetition of the scheme is shown in Table~\ref{M4s3}.
\begin{table}[h]
	\centering
	\caption{Meaningful queries for $M=4$, $N=2$, $s_2=3$.}
	\label{M4s3}
	\begin{tabular}{|c|c|}
		\hline
		Database 1 & Database 2 \\
		\hline
		$a_1,b_1,c_1,d_1$ & $a_2+b_1+c_1+d_1$ \\
		\hline
	\end{tabular}
\end{table}

\paragraph{The scheme corresponding to $\mathbf{n}=(1,1,2,2)$:} In this case $s_2=2$, hence the user combines every $2$ undesired symbols from database 1 to form one side information symbol. To that end, the user downloads $\binom{M-2}{s_2-1}=2$ stages of individual symbols (1-sum) from database 1, so that the user forms 2-sums that can be used in database 2 as side information to start round 3 directly. More specifically, the user downloads $a_3+b_1+c_1$, $a_4+b_2+d_1$, $a_5+c_2+d_2$ from database 2 taking into considerations that all these undesired symbols are decodable from database 1. The user completes the stage by downloading $b_3+c_3+d_3$ that can be further exploited in database 1 by downloading $a_6+b_3+c_3+d_3$. Hence, $D_1(\mathbf{n})=9$, $D_2(\mathbf{n})=4$. Consequently, $t_1(\mathbf{n},\bmu)=\frac{9\nu}{1-\mu_1}$ and $t_2(\mathbf{n},\bmu)=\frac{4\nu}{1-\mu_2}$. As the user decodes $6$ symbols from $W_1$ in each repetition, $L(\mathbf{n})=6\nu$. Hence, $R(\mathbf{n}, \bmu)=\frac{6}{\frac{9}{1-\mu_1}+\frac{4}{1-\mu_2}}$ is achievable. The query table of the meaningful queries (without the artificial noise) for one repetition of the scheme is shown in Table~\ref{M4s2}.
\begin{table}[h]
	\centering
	\caption{Meaningful queries for $M=4$, $N=2$, $s_2=2$.}
	\label{M4s2}
	\begin{tabular}{|c|c|}
		\hline
		Database 1 & Database 2 \\
		\hline
		$a_1,b_1,c_1,d_1$ & $a_3+b_1+c_1$ \\
		$a_2,b_2,c_2,d_2$ & $a_4+b_2+d_1$\\
		                  & $a_5+c_2+d_2$\\
		                  & $b_3+c_3+d_3$\\
		\hline
		$a_6+b_3+c_3+d_3$&\\
		\hline
	\end{tabular}
\end{table}

\paragraph{The scheme corresponding to $\mathbf{n}=(1,2,2,2)$:} In this case $s_2=1$, hence the user exploits the individual undesired symbols downloaded from database 1 directly as a side information in database 2. To that end, the user exploits the side information generated in round 1 by downloading $a_2+b_1$, $a_3+c_1$, and $a_4+d_1$. The user completes the stage by downloading undesired symbols consisting of 2-sums that do not include $a_i$, hence the user downloads $b_2+c_2$, $b_3+d_2$, $c_3+d_3$. The undesired symbols are exploited in database 1, thus the user downloads $a_5+b_2+c_2$, $a_6+b_3+d_2$, and $a_7+c_3+d_3$. The user completes the stage by downloading $b_4+c_4+d_4$, which can be exploited in database 2 by downloading $a_8+b_4+c_4+d_4$. Hence, $D_1(\mathbf{n})=8$, $D_2(\mathbf{n})=7$. Consequently, $t_1(\mathbf{n},\bmu)=\frac{8\nu}{1-\mu_1}$, and $t_2(\mathbf{n},\bmu)=\frac{7\nu}{1-\mu_2}$. As the user decodes $8$ symbols from $W_1$ in each repetition, $L(\mathbf{n})=8\nu$. Hence, $R(\mathbf{n}, \bmu)=\frac{8}{\frac{8}{1-\mu_1}+\frac{7}{1-\mu_2}}$ is achievable. The query table of the meaningful queries (without the artificial noise) for one repetition of the scheme is shown in Table~\ref{M4s1}.

\begin{table}[h]
	\centering
	\caption{The query table for $M=4$, $N=2$, $s_2=1$.}
	\label{M4s1}
	\begin{tabular}{|c|c|}
		\hline
		Database 1 & Database 2 \\
		\hline
		$a_1,b_1,c_1,d_1$ &$a_2+b_1$  \\
					      &$a_3+c_1$  \\
						  &$a_4+d_1$  \\
						  &$b_2+c_2$  \\
						  &$b_3+d_2$  \\
						  &$c_3+d_3$  \\
		\hline
		
		\hline
		$a_5+b_2+c_2$&  $a_8+b_4+c_4+d_4$\\
		$a_6+b_3+d_2$&\\
		$a_7+c_3+d_3$&\\
		$b_4+c_4+d_4$&\\
		\hline
	\end{tabular}
\end{table}
As in the case of $M=3$, under the assumption that $\mu_1 \leq \mu_2$, the symmetric scheme in \cite{JafarPIR} does not achieve any larger retrieval rates at any $\bmu$.
Hence, the following rate is achievable,
\begin{align}
R(\bmu)=\max\left\{\frac{1-\mu_1}{4},\:\frac{2}{\frac{4}{1-\mu_1}+\frac{1}{1-\mu_2}},\:\frac{6}{\frac{9}{1-\mu_1}+\frac{4}{1-\mu_2}},\:\frac{8}{\frac{8}{1-\mu_1}+\frac{7}{1-\mu_2}} \right\}
\end{align}

\begin{figure}[t]
	\centering
	\includegraphics[width=0.8\textwidth]{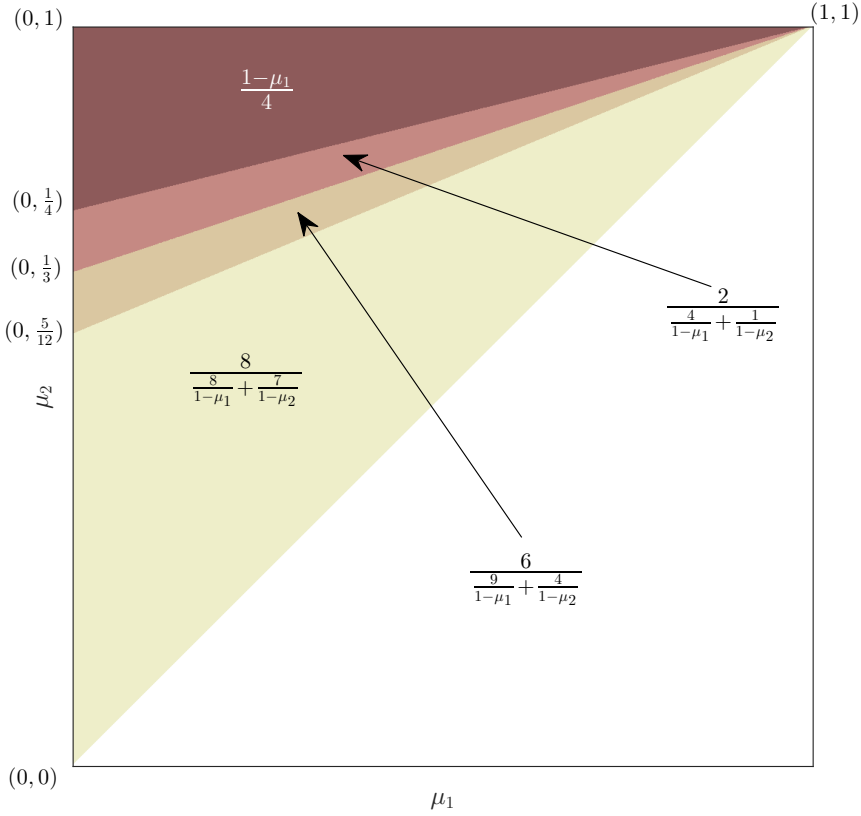}
	\caption{Partitions of $\bmu$ space according to retrieval rate expression for $M=4$, $N=2$.}
	\label{Fig: RegionsM4N2}
	\vspace*{-0.4cm}
\end{figure}

\begin{figure}[t]
	\centering
	\includegraphics[width=0.8\textwidth]{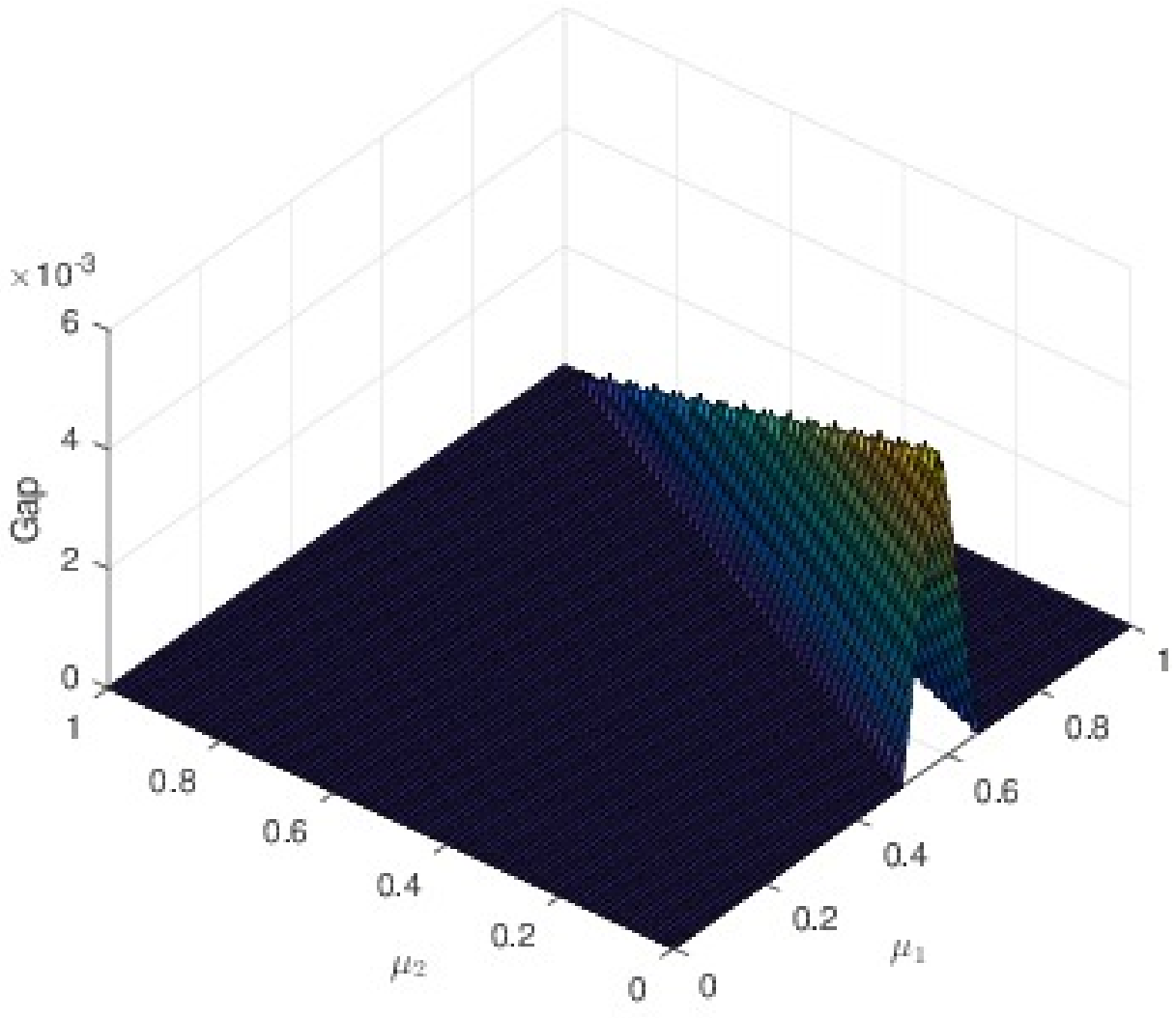}
	\caption{Capacity gap for the case of $M=4$, $N=2$.}
	\label{Fig:gap3D}
	\vspace*{-0.4cm}
\end{figure}

In Fig.~\ref{Fig: RegionsM4N2}, we illustrate the partitioning of the $\bmu$ space in terms of the active achievable scheme. In Fig.~\ref{Fig:gap3D}, we plot the gap versus $\bmu$ for $M=4$, $N=2$. We note that the gap is upper bounded by $0.0051$ and this gap exists only for specific regimes of $\bmu$.

\subsubsection{$M=2$ Messages, $N=3$ Databases}
In this example, we show the achievable scheme for $M=2$, $N=3$, and arbitrary $\bmu$. Again we focus on the meaningful queries in our exposition to avoid repetition. The artificial noise incorporation is exactly as in the previous examples.  Let $a_i, b_i$ denote the randomly permuted symbols from $W_1, W_2$, respectively.

\paragraph{The trivial scheme corresponding to $(n_0,n_1)=(1,1)$:} In this case, the user deactivates the retrieval from database 2. Hence, in one repetition, the user downloads $a_1,b_1$ from database 1 only. Therefore, $D_1(1,1)=2$ which leads to $t_1(1,1,\bmu)=\frac{2\nu}{1-\mu_1}$. From one repetition of the scheme, the user decodes $1$ symbol from $W_1$, hence $L=\nu$ symbols. This gives the rate $R(1,1,\bmu)=\frac{1-\mu_1}{2}$.

\paragraph{The scheme corresponding to $(n_0,n_1)=(1,2)$:}
In this case, the user exploits the undesired symbols in database 1 as a side information in database 2 only and deactivates database 3. Hence, in one repetition, the user downloads $a_1, b_1$ from database 1, and uses $b_1$ as side information in database 2 by downloading $a_2+b_1$. Therefore, $D_1(1,2)=2$, $D_2(1,2)=1$ which leads to $t_1(1,2,\bmu)=\frac{2\nu}{1-\mu_1}$, and $t_2(1,2,\bmu)=\frac{\nu}{1-\mu_2}$. From one repetition of the scheme, the user decodes $2$ symbols from $W_1$, hence $L=2\nu$ symbols. This gives the rate $R(1,2,\bmu)=\frac{2}{\frac{2}{1-\mu_1}+\frac{1}{1-\mu_2}}$. The query table of the meaningful queries (without the artificial noise) for one repetition of the scheme is shown in Table~\ref{M2n12}.
\begin{table}[h]
	\centering
	\caption{Meaningful queries for $M=2$, $N=3$, $\mathbf{n}=(1,2)$.}
	\label{M2n12}
	\begin{tabular}{|c|c|c|}
		\hline
		Database 1 & Database 2 & Database 3 \\
		\hline
		$a_1,b_1$ & $a_2+b_1$ & \\
		\hline
	\end{tabular}
\end{table}
\paragraph{The scheme corresponding to $(n_0,n_1)=(1,3)$:}
Since $n_1=3$, the user exploits the side information in database 2 and database 3. Hence, in one repetition, the user downloads $a_1,b_1$ from database 1. The user downloads $a_2+b_1$ from database 2, and $a_3+b_1$ from database 3. Therefore, $D_1(1,3)=2$, $D_2(1,3)=1$, $D_3(1,3)=1$ which leads to $t_1(1,3,\bmu)=\frac{2\nu}{1-\mu_1}$, $t_2(1,3,\bmu)=\frac{\nu}{1-\mu_2}$, $t_3(1,3,\bmu)=\frac{\nu}{1-\mu_3}$. From one repetition of the scheme, the user decodes $3$ symbols from $W_1$, hence $L=3\nu$ symbols. This corresponds to the rate $R(1,3,\bmu)=\frac{3}{\frac{2}{1-\mu_1}+\frac{1}{1-\mu_2}+\frac{1}{1-\mu_3}}$. The query table of the meaningful queries (without the artificial noise) for one repetition of the scheme is shown in Table~\ref{M2n13}.
\begin{table}[h]
	\centering
	\caption{Meaningful queries for $M=2$, $N=3$, $\mathbf{n}=(1,3)$.}
	\label{M2n13}
	\begin{tabular}{|c|c|c|}
		\hline
		Database 1 & Database 2 & Database 3 \\
		\hline
		$a_1,b_1$ & $a_2+b_1$ & $a_3+b_1$\\
		\hline
	\end{tabular}
\end{table}
\paragraph{The scheme corresponding to $(n_0,n_1)=(2,2)$:}
In this case, the user applies the symmetric scheme at databases 1 and 2, and deactivates database 3. Consequently, the user downloads $a_1,b_1$ from database 1. From database 2, the user downloads new symbols $a_2,b_2$. The user exploits the side information generated in the first round of download by downloading $a_3+b_2$, and $a_4+b_1$. Therefore, $D_1(2,2)=3$, $D_2(2,2)=3$ which leads to $t_1(2,2,\bmu)=\frac{3\nu}{1-\mu_1}$, $t_2(2,2,\bmu)=\frac{3\nu}{1-\mu_2}$. From one repetition of the scheme, the user decodes $4$ symbols from $W_1$, hence $L=4\nu$ symbols. This gives the rate $R(2,2,\bmu)=\frac{4}{\frac{3}{1-\mu_1}+\frac{3}{1-\mu_2}}$. The query table of the meaningful queries (without the artificial noise) for one repetition of the scheme is shown in Table~\ref{M2n22}.
\begin{table}[h]
	\centering
	\caption{Meaningful queries for $M=2$, $N=3$, $\mathbf{n}=(2,2)$.}
	\label{M2n22}
	\begin{tabular}{|c|c|c|}
		\hline
		Database 1 & Database 2 & Database 3 \\
		\hline
		$a_1,b_1$ & $a_2,b_2$ & \\
		\hline
		$a_3+b_2$ & $a_4+b_1$ & \\
		\hline
	\end{tabular}
\end{table}

\paragraph{The scheme corresponding to $(n_0,n_1)=(2,3)$:}
In this case, the user further exploits the side information generated in databases 1 and 2 in database 3. Hence, the user downloads $a_3+b_1$, $a_4+b_2$ from database 3. Therefore, $D_1(2,3)=3$, $D_2(2,3)=3$, $D_3(2,3)=2$ which leads to $t_1(2,3,\bmu)=\frac{3\nu}{1-\mu_1}$, $t_2(2,3,\bmu)=\frac{3\nu}{1-\mu_2}$, $t_3(2,3,\bmu)=\frac{2\nu}{1-\mu_3}$. From one repetition of the scheme, the user decodes $6$ symbols from $W_1$, hence $L=6\nu$ symbols. This gives the rate $R(2,3,\bmu)=\frac{6}{\frac{3}{1-\mu_1}+\frac{3}{1-\mu_2}+\frac{2}{1-\mu_3}}$. The query table of the meaningful queries (without the artificial noise) for one repetition of the scheme is shown in Table~\ref{M2n23}.
\begin{table}[h]
	\centering
	\caption{Meaningful queries for $M=2$, $N=3$, $\mathbf{n}=(2,3)$.}
	\label{M2n23}
	\begin{tabular}{|c|c|c|}
		\hline
		Database 1 & Database 2 & Database 3 \\
		\hline
		$a_1,b_1$ & $a_2,b_2$ & $a_3+b_1$\\
		          &           & $a_4+b_2$\\
		\hline
		$a_5+b_2$ & $a_6+b_1$ & \\
		\hline
	\end{tabular}
\end{table}

\paragraph{The scheme corresponding to $(n_0,n_1)=(3,3)$:} In this case, the user applies the symmetric scheme in \cite{JafarPIR}. 
Therefore, $D_n(3,3)=4$, where $n=1,2,3$ which leads to $t_n(3,3,\bmu)=\frac{4\nu}{1-\mu_n}$. From one repetition of the scheme, the user decodes $9$ symbols from $W_1$, hence $L=9\nu$ symbols. This gives the rate $R(3,3,\bmu)=\frac{9}{\frac{4}{1-\mu_1}+\frac{4}{1-\mu_2}+\frac{4}{1-\mu_3}}$. The query table of the meaningful queries (without the artificial noise) for one repetition of the scheme is shown in Table~\ref{M2n33}.
\begin{table}[h]
	\centering
	\caption{Meaningful queries for $M=2$, $N=3$, $\mathbf{n}=(3,3)$.}
	\label{M2n33}
	\begin{tabular}{|c|c|c|}
		\hline
		Database 1 & Database 2 & Database 3 \\
		\hline
		$a_1,b_1$ & $a_2,b_2$ & $a_3,b_3$ \\
		\hline
		$a_4+b_2$ & $a_6+b_1$ & $a_8+b_1$ \\
		$a_5+b_3$ & $a_7+b_3$ & $a_9+b_2$\\
		\hline
	\end{tabular}
\end{table}

Consequently, the following rate is achievable:
\begin{align}
R(\bmu)=&\max\left\{\frac{1-\mu_1}{2},\: \frac{2}{\frac{2}{1-\mu_1}+\frac{1}{1-\mu_2}}, \: \frac{3}{\frac{2}{1-\mu_1}+\frac{1}{1-\mu_2}+\frac{1}{1-\mu_3}},\right.\notag\\ &\qquad\:\left.\quad\frac{4}{\frac{3}{1-\mu_1}+\frac{3}{1-\mu_2}}, \:
\frac{6}{\frac{3}{1-\mu_1}+\frac{3}{1-\mu_2}+\frac{2}{1-\mu_3}},\: \frac{9}{\frac{4}{1-\mu_1}+\frac{4}{1-\mu_2}+\frac{4}{1-\mu_3}}\right\}
\end{align}

\section{Conclusion}
In this paper, we investigated the PIR-WTC-II problem. We have shown that the problem is a concrete example of the PIR problem under asymmetric traffic constraints. We obtained a general upper bound that extends the converse techniques in \cite{KarimAsymmetricPIR}. The converse proof takes the form of a max-min optimization problem. The inner minimization problem derives the tightest upper bound for the retrieval rate for an arbitrary traffic ratio vector $\bt$, while the outer maximization problem optimizes over $\bt$. The core of the achievability proof is the achievability proof of the corner points of the PIR problem under asymmetric traffic constraints. The security constraint is satisfied by encrypting each returned answering string by an artificial noise vector. To generate the artificial noise vector, the $n$th database generates a secret key and encodes it into artificial noise by a $(t_n,\mu_n t_n)$ MDS code. The upper and lower bounds match for $M=2$ and $M=3$, for any $N$, and for every eavesdropping capability vector $\bmu=(\mu_1, \cdots, \mu_N)$.

\bibliographystyle{unsrt}
\bibliography{references}
\end{document}